# The Quantum Theory of the Lorentzian Fermionic Differential Forms

A. Jourjine

FG CTP
Hofmann Str. 6-8
01277 Dresden
Germany

**Abstract**

We consider the quantum theory of the Lorentzian fermionic differential forms and the corresponding bi-spinor quantum fields, which are expansion coefficients of the forms in the bi-spinor basis of Becher and Joos [7]. The canonical quantization procedure for the bi-spinor gauge theory in terms of its Dirac spinor constituents is described in detail and the corresponding Feynman rules are derived. We also derive all possible mass terms for massive fermions in the bi-spinor gauge theory. The solutions are classified by a scalar spin quantum number, a number that has no analog in the standard gauge theory and in the SM. The possible mass terms correspond to combinations of scalar spin zero singlets and scalar spin one-half doublets in the generation space. A description of the connection between Lorentz spin of bi-spinors and scalar spin of bi-spinor constituents is given.

## 1. Introduction

The notion of the differential form and its use in physics as a descriptor of fermionic matter is as old as that of the spinor. It was first used by Ivanenko and Landau in 1928 [1], the same year Dirac proposed his theory of the electron [2]. Using antisymmetric tensors that describe a general differential form in the coordinate basis Ivanenko and Landau constructed an alternative to Dirac's solution of the electron's gyromagnetic ratio problem. The coefficients of a general differential form in another basis that transforms in the bi-fundamental representation of the Lorentz group, we call them bi-spinors, are the main mathematical objects in our work as the bi-spinor quantized fields. The mathematical setting for bi-spinors in terms of differential forms on space-time manifolds was described in [3] and further elucidated in [4]. In the 80's it was found that a single bi-spinor describes multiple generations of particles [5, 6, 7, 8].

One of the interesting properties of bi-spinors is that in the presence of gravitational fields they are physically distinct from Dirac spinors [5, 6, 8]. It turned out that the Einstein-Hilbert gravity theory, which cannot describe Dirac spinors[1], incorporates bi-spinors in an elegant form. It was also noticed that on pseudo-Euclidean space-times there exist difficulties with the quantization of bi-spinors using the standard Dirac quantization procedure [7, 9, 10]. A proposal to eliminate the unwanted modes using indefinite Hilbert norm along the lines of Gupta-Bleuler formalism was not successful [11].

Bi-spinors on the lattice turned out to be unavoidable. They appear as a result of fermion doubling effect and there they describe multiplets of particles as well. They also were studied in lattice supersymmetry [12, 13, 14, 15, 16]. Anti-symmetric tensor forms of bi-spinors appear often in string theories as bosonic p-forms. Their quantization has been studied both in supergravity and in string theory [17, 18, 19, 20].

[1] Dirac spinors require the first order, for example, the Cartan formulation of gravity where frames are coupled to spin connection.

Recently it has been shown that the unwanted features of bi-spinor gauge theories – the unobserved 4$^{th}$ generation and the unwanted quantization modes - may be avoided. The fourth generation, which appears in the bi-spinor field theory generically, can be eliminated with a help of a natural constraint [22, 23]. As for the second major obstacle, as we will show in this paper, the quantization problem can be resolved by modifying the standard Dirac quantization rules for some of the bi-spinor modes.

With the two major obstacles removed, bi-spinor gauge theories may be considered as a viable alternative to the Standard Model. This paper lays a foundation for such a bi-spinor Standard Model by detailed analysis of the relevant bi-spinor gauge theory and its perturbation theory. Despite the progress in understanding the nature of bi-spinor theories, a consistent quantum field theory of bi-spinors and the corresponding perturbation theory of its Dirac degrees of freedom have never been constructed. A serious attempt in [10] was made of carrying out the canonical quantization of bi-spinors in terms of quantum amplitudes transforming as bi-spinors. It resulted, as expected, in the need to have two Lorentz spin quantum numbers, the left and the right spin. However, the results were not transparent physically, since particles with two Lorentz spins are not observed.

In this paper we avoid this problem by first extracting the Dirac degrees of freedom from bi-spinors and then quantizing the Dirac modes only. Extending the results in [22, 23], we will show that on Minkowski space-times bi-spinors can be consistently quantized and their formal perturbation theory can be built along the standard prescriptions of the standard quantum field theory. On the way to this goal we will exhibit a number of curious features of quantum field theory of bi-spinors. It differs from the Dirac theory in a number of subtle but fundamental ways which offer possible solutions for some long-standing puzzles of the elementary particle physics, such as the textures of the CKM and PMNS flavor mixing matrices. The physical reason for the differences is the existence in the bi-spinor gauge theory of a new quantum number [21], called scalar spin, which is inherited by the Dirac degrees of freedom from the bi-spinor nature of bi-spinors.

Since in the literature the coefficients of the fermionic[2] inhomogeneous differential forms are often referred either as Dirac-Kähler or Ivanenko-Landau spinors, a few words are in order to explain our choice of the terminology. We use the term bi-spinor for the same reason the term gauge vector field is used to denote a bosonic differential 1-form. The vector potential is a coefficient of decomposition of the invariant gauge field 1-form in the coordinate basis. Following the same logic we use the term bi-spinor, since bi-spinors are the coefficients of the general fermionic differential form in the bi-spinor basis. Unlike coefficients in the coordinate basis they transform in the bi-fundamental representation of the Lorentz group. We should also point out that the common belief that a bi-spinor field is a collection of four Dirac spinors is justified only for Euclidean space-times. As we will see below, for Lorentzian space-times such description is inadequate, because in addition to four fermionic Dirac spinors a Lorentzian bi-spinor also contains four bosonic Dirac spinors, whose presence is essential, for it determines the masses of the particles.

A differential form can be expanded in the coordinate basis, resulting in a collection of anti-symmetric tensors as expansion coefficients, or in the bi-spinor basis, resulting in a bi-spinor coefficient matrix. Algebraically both forms are equivalent. However, there is a fundamental difference on the level of dynamics. It is the dynamics that determines the choice of the appropriate basis: if the differential forms are commuting we need to use the coordinate basis to obtain integer spin equations of motion for the fields. If, however, the differential forms are anti-commuting one has to use the bi-spinor basis to obtain the spin one-half equations of motion for the fields. Thus the statistics of the elementary particles determines the expansion basis. As a result, a collection of antisymmetric tensor fields is not appropriate for description of fermionic degrees of freedom. Such a collection is best suited for description of extended gauge theories [24, 25].

[2] By fermionic/bosonic differential form we understand the one whose quantization requires the use of anti/commutators. The same terminology is carried over to the basis expansion coefficients.

The paper is organized as follows. In Section 2 we recapitulate the results of [22, 23] on the extraction of Dirac degrees of freedom from bi-spinors. In Section 3 we derive a classification of all possible mass terms in bi-spinor gauge theories. Section 4 describes the canonical quantization of Dirac and anti-Dirac spinor singlets. The quantization of the Dirac-anti-Dirac (DaD) spinor doublets is described in Section 5. In Section 6 we describe the connection between the Lorentz spin of the bi-spinors and the scalar spin of its Dirac/anti-Dirac constituents. Section 7 is a summary. Appendix A contains details of derivation of the plane wave solutions for DaD spinor doublets. Appendix B describes the diagonalization of the DaD doublet Hamiltonian. Appendix C contains a discussion of the scalar spin states in Fock space. Appendix D lists the Feynman rules for bi-spinors. Below we follow the conventions of [26]. We work with natural units where all fundamental constants are set to one.

## 2. Extraction of Algebraic Dirac Spinors from Bi-Spinors via Spinbeins

Consider Minkowski space-time $M_4$ with coordinates $x^\mu$, $\mu = 0,\dots,3$, metric $g_{\mu\nu} = diag(+1,-1,-1,-1)$ and a set of Dirac $\gamma$-matrices $\{\gamma^\mu, \gamma^\nu\} = 2g^{\mu\nu}$. In the bi-spinor gauge theory bosonic gauge fields are described by the usual connection 1-forms. In the coordinate basis (c-basis) they are given by

$$A = A_\mu dx^\mu, \qquad A_\varpi = A_\mu^a \tau^a, \quad tr(\tau^a \tau^b) = \frac{1}{2}\delta^{ab}, \qquad \tau^a = 1 \quad \text{for } U(1), \tag{2.1}$$

where $\tau^a$, $a = 1,\dots,N_G$, are the $N_G$ generators of the Lie algebra of gauge group $G$. The fermionic degrees of freedom are described by anti-commuting inhomogeneous differential forms (difforms). In the c-basis they can be represented by

$$\Phi = \sum_p \Phi_p = \sum_p \frac{1}{p!} \Phi_{\mu_1\dots\mu_p} dx^{\mu_1} \wedge \dots \wedge dx^{\mu_p}. \tag{2.2}$$

Difforms $\Phi$ are invariant objects that actually do not depend on the system of coordinates. By choosing another basis, a certain z-basis, in the space of difforms a difform $\Phi$ can be equivalently represented in another way, as a bi-spinor $\Psi(\Phi) = \{\Psi_{\alpha\beta}(\Phi)\}$. Therefore, one can use both a set of anti-symmetric tensors $\Phi_{\mu_1\dots\mu_p}$ and a bi-spinor $\Psi(\Phi)$, which transforms in the adjoint spinor representation of $SO(1,3)$: $\Psi(\Phi) \to S(\Lambda)\Psi(\Phi)S^{-1}(\Lambda)$, to represent the same mathematical object. Relations between difform $\Phi$, and the two sets of coefficients of its expansion, $\Phi_{\mu_1\dots\mu_p}$ in the c-basis and $\Psi(\Phi)$ in the z-basis, follow from the completeness relations of gamma-matrices, with the use of a $4\times4$ matrix of difforms $Z$ [7] given by

$$(Z)_{\alpha\beta} = \sum_p \frac{1}{p!} \left(\gamma_{\mu_p} \cdots \gamma_{\mu_1}\right)_{\alpha\beta} dx^{\mu_1} \wedge \dots \wedge dx^{\mu_p}.$$

One obtains

$$\Phi = tr\left(Z\,\Psi(\Phi)\right),$$

$$\Psi(\Phi) = \frac{1}{4}\sum \frac{1}{p!}\gamma^{\mu_1}\cdots\gamma^{\mu_p}\Phi_{\mu_1\cdots\mu_p}\,, \tag{2.3}$$

$$\Phi_{\mu_1\cdots\mu_p} = tr\left(\gamma_{\mu_p}\cdots\gamma_{\mu_1}\,\Psi(\Phi)\right),$$

where trace is taken over spinor indices.

Bi-spinor gauge theory can be expressed in the invariant language of difforms on a smooth manifold without a reference to either c- or z-basis. However, since our emphasis in this paper is on quantization of bi-spinor gauge theory and derivation of its Feynman rules, we will not concern ourselves further with the connection between differential geometry and bi-spinors.

A bi-spinor gauge theory with massive bi-spinors, where left- and right-handed fermions could couple to different gauge groups $G_L, G_R$, to be concrete we can use $G_L = SU(3)\times SU(2)\times U(1)$ and $G_R = SU(3)\times U(1)$, is described by the Lagrangian density [22, 23]

$$\mathcal{L} = \mathcal{L}_g + \mathcal{L}_f\,, \tag{2.4}$$

$$\mathcal{L}_g = -\frac{1}{4}B_{\mu\nu}B^{\mu\nu} - \frac{1}{2}tr W_{\mu\nu}W^{\mu\nu} - \frac{1}{2}tr\, G_{\mu\nu}G^{\mu\nu}, \tag{2.5}$$

$$\mathcal{L}_f = tr\left[\overline{\overline{\Psi}}_L\left(i\partial\!\!\!/ + A\!\!\!/_L\right)\Psi_L + \overline{\overline{\Psi}}_R\left(i\partial\!\!\!/ + A\!\!\!/_R\right)\Psi_R - m\left(\overline{\overline{\Psi}}_L\,\Psi_R\,M + \overline{\overline{\Psi}}_R\,\Psi_L\,\tilde{M}\right)\right], \tag{2.6}$$

$$\tilde{M} = \gamma^0 M^+ \gamma^0, \qquad \overline{\overline{\Psi}}_{L,R} = \gamma^0 \Psi_{L,R}{}^+ \gamma^0\,, \tag{2.7}$$

where in our example $B_{\mu\nu}, W_{\mu\nu}, G_{\mu\nu}$ are the field strength of the irreducible components of the left-handed connection 1-form $A_L = \left(gB_\mu + g' W_\mu{}^a \tau^a + g_s G_\mu{}^s T^s\right)dx^\mu$ and right-handed connection 1-form $A_R = \left(gB_\mu + g_s G_\mu{}^s T^s\right)dx^\mu$ and $\tau^a, T^s$ are Lie algebra generators of $SU(2)$ and $SU(3)$ factors, respectively.

The explicit mass term is given by a constant matrix $mM$. For convenience we chose $M$ dimensionless. At this point it is an arbitrary complex matrix. As we will see in Section 3, $M$ is actually severely restricted.

In order for the mass term be gauge-invariant $\Psi_L$ and $\Psi_R$ have to transform in the same representation of the gauge group. In the Dirac gauge theory if $G_L \neq G_R$ then there are no explicit gauge-invariant mass terms. To generate mass of the particles one has to use an additional field, the Higgs field, the vacuum expectation value of which generates a mass-like term in (2.6).

In bi-spinor gauge theory, although Higgs fields coupled to bi-spinors remain allowed, explicit gauge-invariant mass terms are also allowed. The reason for this is that the gauge group representation in which bi-spinors transform is the bi-fundamental representation. It is a direct product of the not necessarily the same fundamental representations in which the constituents in bi-spinor spinbein decomposition transform. Therefore, for example, one can combine the left-handed Dirac spinor $SU(2)$ doublet with a right-handed Dirac spinor $SU(2)$ singlet in a gauge-invariant mass term, provided the bi-fundamental representations are chosen appropriately [23].

Equations of motion for $\Psi_{L,R}$ can be read from (2.6). We obtain

$$\left(i\partial\!\!\!/ + A\!\!\!/_L\right)\Psi_L - m\Psi_R M = 0\,, \tag{2.8}$$

$$\left(i\partial\!\!\!/ + A\!\!\!/_R\right)\Psi_R - m\Psi_L \tilde{M} = 0\,. \tag{2.9}$$

First we will consider in detail the simplest $U(1)$ case and then follow with a discussion of the non-Abelian case. Mass eigenstates of $U(1)$ bi-spinors are determined by free Lagrangian density and the corresponding equations of motion

$$\mathcal{L}_0 = tr\left[\overline{\overline{\Psi}}_L (i\partial\!\!\!/)\Psi_L + \overline{\overline{\Psi}}_R (i\partial\!\!\!/)\Psi_R - m\left(\overline{\overline{\Psi}}_L\,\Psi_R M + \overline{\overline{\Psi}}_R\,\Psi_L \tilde{M}\right)\right], \tag{2.10}$$

$$i\partial\!\!\!/\Psi_L - m\,\Psi_R M = 0\,, \tag{2.11}$$

$$i\partial\!\!\!/\Psi_R - m\Psi_L \tilde{M} = 0\,. \tag{2.12}$$

We will now derive the Lagrangian and equations of motion for the Dirac spinor degrees of freedom that are contained in bi-spinors $\Psi_{L,R}$ in (2.6). To start with, we will consider bi-spinors that contain four generations of Dirac spinors. Reduction to less then four generations will be considered at the end of this section. First, we will extract the algebraic Dirac spinors that contain the physical degrees of freedom using the spinbein decomposition of bi-spinors. This is done with the help of two sets of algebraic Dirac spinors: the anti-commuting Dirac spinors $\xi^A$, $A = 1,\ldots,4$, and the spinbein, a set of four dimensionless normalized commuting Dirac spinors $\eta^A$, $A = 1,\ldots,4$. Eventually we will promote $\xi^A$ to quantum fields, leaving $\eta^A$ as classical, commuting objects. The spinbein is normalized (note that unless explicitly noted we always understand summation over repeated indices) by setting

$$\begin{gathered}\overline{\overline{\eta}}{}^A_\alpha\,\eta^B_\alpha = \delta^{AB}\,,\\ \overline{\overline{\eta}}{}^A{}_\alpha = \Gamma^{AB}\eta^{+B}{}_\beta\,\gamma^0{}_{\beta\alpha}\,, \qquad \overline{\overline{\eta}} = \Gamma\eta^+\gamma^0 \quad \Gamma = diag\left(1,\ 1,-1,-1\right),\end{gathered} \tag{2.13}$$

where $\overline{\eta} = \eta^+\gamma^0$ denotes the Dirac conjugate and $\overline{\overline{\eta}}{}^A$ denotes the bi-spinor conjugate of $\eta^A$. We will refer to spinbeins that carry gauge group indices as gauged spinbeins. $U(1)$ or non-gauged spinbeins carry no representation indices and we can invert (2.13) to obtain

$$\eta^A_\alpha \overline{\overline{\eta}}{}^A_\beta = \delta_{\alpha\beta}\,. \tag{2.14}$$

For a spinbein $\left\{\eta^A\right\}$ and a multiplet of four Dirac spinors $\xi^A$ spinbein decomposition is defined by

$$\Psi = \xi\,\overline{\overline{\eta}}\,, \qquad \overline{\overline{\Psi}} = \eta\,\overline{\overline{\xi}}\,, \qquad \Psi_{\alpha\beta} = \xi^A_\alpha\,\overline{\overline{\eta}}{}^A_\beta\,, \qquad \overline{\overline{\Psi}}_{\alpha\beta} = \eta^A_\alpha\overline{\overline{\xi}}{}^A_\beta\,. \tag{2.15}$$

We will assume that the left- and the right-handed bi-spinors $\Psi_{L,R}(x)$ are independent dynamical variables. As a result, we will have to use two separate spinbeins for their decomposition

$$\Psi_L(x)=\psi_L(x)\overline{\overline{\eta}}_L(x), \qquad \Psi_R(x)=\psi_R(x)\overline{\overline{\eta}}_R(x), \qquad (1\pm\gamma^5)\psi_{L,R}=0. \tag{2.16}$$

Note that $\eta_{L,R}(x)$ are not chiral: $(1\pm\gamma^5)\eta_{L,R}\neq 0$.

In the Dirac representation of gamma-matrices with $\gamma^0 = diag(1,1,-1,-1)\equiv\Gamma$ spinbeins can be identified with the elements of $U(2,2)$, the group of all complex $4\times 4$ matrices that satisfy

$$\overline{\overline{U}}U=1, \qquad \overline{\overline{U}}\equiv\Gamma U^{+}\Gamma. \tag{2.17}$$

A well-known example of spinbein can be found in the phase space of Dirac spinors satisfying Dirac equation. It appears during the derivation of normalized plane wave solutions of the Dirac equation on $M_4$ as spinor coefficients of the four positive and negative energy solutions $u^r(k)\exp(-ikx),\ v^s(k)\exp(+ikx),\ r,s=1,2$. The spinors $u^r(k),\ v^s(k)$ are normalized according to

$$\bar{u}^r u^s=\delta^{rs}, \qquad \bar{v}^r v^s=-\delta^{rs}, \qquad \bar{u}^r v^s=0. \tag{2.18}$$

In multi-index notation $w^A=(u^r,v^s)$ with index $A$, $A=(r,s)=1,\ldots,4$, (2.18) reduces to the defining relation (2.17) for $U(2,2)$ group: $\overline{\overline{w}}{}^A_\alpha w^B_\alpha=\delta^{AB}$.

It follows from (2.15) that bi-spinors $\Psi_{L,R}(x)$ are invariant with respect to two independent local $U(2,2)$ transformations. Namely, $\Psi_{L,R}(x)\to\Psi_{L,R}(x)$ if

$$\begin{aligned}
&\psi_L(x)\to\psi_L{}'=\psi_L(x)U_L(x), \quad \eta_L(x)\to\eta_L{}'=\eta_L(x)U_L(x), \quad U_L(x)\in U(2,2).\\
&\psi_R(x)\to\psi_R{}'=\psi_R(x)U_R(x), \quad \eta_R(x)\to\eta_R{}'=\eta_R(x)U_R(x), \quad U_r(x)\in U(2,2).
\end{aligned} \tag{2.19}$$

Therefore, by using spinbein decomposition (2.15) we introduced redundant degrees of freedom. To eliminate the redundancy we will transfer the spinbein degrees of freedom from spinbein to the multiplet of algebraic Dirac spinors $\psi_{L,R}{}^A$ by requiring that the two spinbeins are constant

$$\partial_\mu\eta_L(x)=0, \qquad \partial_\mu\eta_R(x)=0. \tag{2.20}$$

For $U(1)$ the transfer can always be done by using local $U(2,2)$ transformations (2.19), such that $\partial_\mu\eta_{L,R}{}'=0$. Fixing spinbein to be constant is in fact fixing a gauge of a $U(2,2)$ gauge field [22]. Such a gauge is called the unitary or constant gauge. Constant gauge is not unique. There still remains a global $U(2,2)$ symmetry of $\Psi$.

For non-Abelian case with bi-spinor transforming in a bi-fundamental representation the spinbein decomposition is defined by

$$\Psi = \xi\, \overline{\overline{\eta}}\,, \qquad \Psi_{\alpha\beta}{}^{ap} = \xi_\alpha^{aA} \overline{\overline{\eta}}_\beta{}^{pA}, \qquad \overline{\overline{\Psi}}_{\alpha\beta}{}^{pa} = \eta_\alpha^{pA} \overline{\overline{\xi}}_\beta{}^{aA},$$

$$\overline{\overline{\eta}}{}^{sA} = \Gamma^{AB} \overline{\eta}{}^{sB}, \qquad \Gamma = diag(\ 1,\ 1,-1,-1), \tag{2.21}$$

where Dirac spinors and gauged spinbein can transform in fundamental representations of different groups different, while spinbein is normalized according to

$$\overline{\overline{\eta}}_\alpha{}^{sA}\, \eta_\alpha^{sB} = \delta^{AB}. \tag{2.22}$$

We can invert (2.23) only up to a projector

$$\eta_\alpha^{sA} \overline{\overline{\eta}}_\beta^{tA} = P^{st}{}_{\alpha\beta}, \qquad P^2 = P, \qquad \left(P^2\right)^{st}{}_{\alpha\beta} = P^{su}{}_{\alpha\gamma} P^{ut}{}_{\gamma\beta}. \tag{2.23}$$

For non-Abelian gauge groups, in general, it is no longer possible to transfer all degrees of freedom from a gauged spinbein to the multiplet of Dirac spinors. One exception is the case when spinbein factorizes according to

$$\eta_\alpha^{sA} = \phi^s\, \eta_\alpha^A, \tag{2.24}$$

where $\eta$ is a non-gauged spinbein normalized according to (2.13) and Lorentz scalar $\varphi_s$ transforms in the fundamental representation of the gauge group. Normalization of the gauged spinbein implies that its factors satisfy

$$\phi^{s^*} \phi^s = 1,$$

$$\eta_\alpha^A\, \overline{\overline{\eta}}_\beta{}^A = \delta_{\alpha\beta}, \tag{2.25}$$

$$\eta_\alpha^{sA} \overline{\overline{\eta}}_\beta^{tA} = \varphi^s \varphi^{t^*} \delta_{\alpha\beta}.$$

It turns out that, just like ungauged spinbeins, factorizable spinbeins are non-dynamical [23]. Thus the transfer of degrees of freedom for such spinbeins is also complete. Below we will consider only factorizable spinbeins.

Equations (2.20) are not generally covariant. Therefore, on curved space-times the constant gauge depends on the choice of coordinates. Since constant gauge depends on the choice of coordinates, Dirac spinors $\psi^A$ extracted from bi-spinors also depend on the choice of coordinates. This means that the definition of physical one-particle states of the fermions described by bi-spinors depends on reference frame. On $M_4$ two constant ungauged spinbeins $\eta_1$, $\eta_2$ are connected by a $U(2,2)$ transformation

$$\eta_1 = \eta_2 U\ , \qquad U \in U(2,2), \tag{2.26}$$

for some constant $U$. For gauged spinbeins (2.26) is augmented by an additional gauge group transformation.

Because they define specific physical particle states, spinbeins are physical quantities. Yet at the same time they are not observable, because they are non-dynamical. Their role is similar to the role played by the constant magnetic field that determines the preferred magnetization in a ferromagnetic material.

We now turn to equations of motion for Dirac spinor constituents of bi-spinors. Using spinbein decomposition with constant spinbein we obtain the Lagrangian and equations of motion for Dirac degrees of freedom of bi-spinors

$$\mathcal{L} = tr\left[\overline{\overline{\psi}}{}_L^A\left(i\partial\!\!\!/ + A\!\!\!/_L\right)\psi_L^A + \overline{\overline{\psi}}{}_R^A\left(i\partial\!\!\!/ + A\!\!\!/_R\right)\psi_R^A - m\left(\overline{\overline{\psi}}{}_L^A\,\mathcal{M}^{AB}\psi_R^B + \overline{\overline{\psi}}{}_R^A\,\tilde{\mathcal{M}}^{AB}\psi_L^B\right)\right], \tag{2.27}$$

$$\begin{aligned} &\left(i\partial\!\!\!/ + A\!\!\!/_L\right)\psi_L^A - m\,\mathcal{M}^{AB}\psi_R^B = 0\,, \\ &\left(i\partial\!\!\!/ + A\!\!\!/_R\right)\psi_R^A - m\,\tilde{\mathcal{M}}^{AB}\psi_L^B = 0\,, \end{aligned} \tag{2.28}$$

where for the $U(1)$ case

$$\mathcal{M}^{AB} = \overline{\overline{\eta}}_R{}^B M\,\eta_L^A\;, \qquad \tilde{\mathcal{M}}^{AB} = \overline{\overline{\eta}}_L{}^B \tilde{M}\,\eta_R^A\,, \qquad \overline{\overline{\psi}}{}_{L,R}^A = \Gamma^{AB}\overline{\psi}{}_{L,R}^A\,. \tag{2.29}$$

For general bi-fundamental bi-spinor representation there are two ways to construct gauge-invariant mass terms. To obtain explicit gauge-invariant mass terms, one can use left/right spinbein decompositions

$$\Psi_L = \xi_L\,\overline{\overline{\eta}}_R\,, \qquad \Psi_R = \xi_R\,\overline{\overline{\eta}}_L\;, \tag{2.30}$$

or

$$\Psi_L = \xi_L\,\overline{\overline{\eta}}_L\,, \qquad \Psi_R = \xi_R\,\overline{\overline{\eta}}_R\;, \tag{2.31}$$

where for the spinbeins subscript left/rights refers to the gauge group for left/right fermions. We now can form gauge invariant terms using appropriate contractions. We obtain for the mass term

$$\mathcal{L}_m = -m\,tr\left(\overline{\overline{\psi}}{}_L^{aA}\,\mathcal{M}^{(ap)AB}\psi_R^{pB} + c.c.\right), \tag{2.32}$$

$$\mathcal{M}^{(ap)AB} = \overline{\overline{\eta}}_R{}^{Bp} M\,\eta_L^{Aa} \qquad \text{or} \qquad \mathcal{M}^{(ap)AB} = \overline{\overline{\eta}}_L{}^{Ba} M\,\eta_R^{Ap}\;, \tag{2.33}$$

where for factorizable spinbeins $\eta_L{}^a = \varphi_R{}^a\eta_L$, $\eta_R{}^a = \varphi_L{}^a\eta_R$ for (2.30) or $\eta_L{}^a = \varphi_R{}^a\eta_L$, $\eta_R{}^a = \varphi_L{}^a\eta_R$ for (2.31) we obtain

$$\mathcal{L}_m = -m\,tr\left(\left(\overline{\overline{\psi}}{}_L^{aA}\cdot\varphi_L{}^a\right)\mathcal{M}^{AB}\left(\psi_R^{pB}\cdot\varphi_R{}^{p^*}\right) + c.c.\right). \tag{2.34}$$

For the SM with $G_L = SU(2)$, $G_R = 1$ where the doublet $\varphi_L{}^C = i\sigma_2\varphi_L{}^*$ transforms in the representation that is equivalent to fundamental representation, after absorption of arbitrary parameter $m$ into the constant doublets $\varphi_L{}^a$, $\varphi_L{}^{a^*}$ we obtain an additional mass term with $\varphi_L{}^C$. Altogether we obtain the standard SM expression for mass term after spontaneous symmetry breaking with a left $SU(2)$ doublet $H_L{}^a$

$$\mathcal{L}_m = -\,tr\left(\left(\overline{\overline{\psi}}{}_L^{aA}\cdot H_L{}^a\right)\mathcal{M}^{AB}\psi_R^B + \left(\overline{\overline{\psi}}{}_L^{aA}\cdot\left((i\sigma_2)H_L{}^*\right)^a\right)\mathcal{M}^{AB}\psi_R^B + c.c.\right),$$

$$H_L{}^a = m\varphi_L{}^a \,. \tag{2.35}$$

We will now describe the elimination of one or more generations from the original four. By construction, a single bi-spinor produces four generations of dynamical spinors: two spinors, corresponding to the first two positive entries in $\Gamma = diag(1, 1, -1, -1)$, and two spinors, corresponding to the negative entries. Reduction from four to less than four generations of $\xi^A$ can be done using generally covariant constraint

$$\det \Psi = 0 \,, \tag{2.36}$$

which for Minkowski space-time can be satisfied in Lorentz-invariant way by the use of a degenerate spinbein. For example, instead of (2.13) or (2.22) one can take

$$\bar{\bar{\eta}}_\alpha^A \, \eta_\alpha^B = diag(1, 1, 1, 0), \qquad \bar{\bar{\eta}}_\alpha^{sA} \, \eta_\alpha^{sB} = diag(1, 1, 1, 0), \tag{2.37}$$

respectively.

One can argue that elimination of an extra generation with the help of a constraint is arbitrary. In certain sense it is. However, it is no more arbitrary then adding two more generations to the first one. In fact it is less arbitrary, because formal addition of two generations leaves no room for posing the question why three. Removing one generation of the four, on the other hand, demands a physical explanation for the reduction.

Note that in the free part of Lagrangian (2.27) Weyl spinors $\psi_{L,R}^A$ for $A = 3,4$ contribute to the action derived from $\mathcal{L}$ with the negative sign of the usual Dirac spinors. The presence of the sign results in important consequences. We will discuss these below. For now we will simply distinguish between $\psi_{L,R}^A$ for $A = 1,2$ and $\psi_{L,R}^A$ for $A = 3,4$ by retaining for the former the name of Dirac (Weyl) spinors but we will call the latter anti-Dirac (anti-Weyl) spinors. We will now proceed with the determination of the mass terms that are physically acceptable.

## 3. Admissible Bi-Spinor Mass Terms

In this section we will prove that the requirement that masses of mass eigenstates in bi-spinor gauge theory are physical puts strong constraints on the form of mass matrices. We will provide a detailed proof for the $U(1)$ case. However, it follows from (2.34) that the results apply to the non-Abelian case for factorizable gauged spinbeins.

Consider equations of motion for free (anti)-Weyl fields $\psi_{L,R}(x)$ in matrix notation

$$\begin{aligned} &(i\partial\!\!\!/)\psi_L - m\mathcal{M}\psi_R = 0\,, && \mathcal{M} = \left(\bar{\bar{\eta}}_R\, M\, \eta_L\right)^T, \\ &(i\partial\!\!\!/)\psi_R - m\tilde{\mathcal{M}}\psi_L = 0\,, && \tilde{\mathcal{M}} = \left(\bar{\bar{\eta}}_L \gamma^0 M^+ \gamma^0\, \eta_R\right)^T = \Gamma\mathcal{M}^+\Gamma\,, \end{aligned} \tag{3.1}$$

where $(...)^T$ means transposition in generation indices. Matrices $\mathcal{M}$, $\tilde{\mathcal{M}}$ are two $4\times 4$ complex matrices with generation indices $A, B = 1,2,3,4$ parameterized by a single $4\times 4$ complex matrix $M = M_{\alpha\beta}$ in (2.10) with spinor indices $\alpha, \beta = 1,2,3,4$. We will use Dirac gamma-matrix representation with diagonal $\gamma^0$. Our results, however, will not depend on the choice of representation.

For plane wave solutions $\psi_{L,R}(x) = \psi^0_{L,R}\, e^{-ikx}$ we obtain the dispersion relations for the left and the right modes as solutions of

$$\det\left(k^2 - m^2 \mathcal{M}\tilde{\mathcal{M}}\right) = 0 \ , \qquad \mathcal{M}\tilde{\mathcal{M}} = \left(\bar{\bar{\eta}}_L \gamma^0 M^+ \gamma^0 M\, \eta_L\right)^T ,$$
$$\det\left(k^2 - m^2 \tilde{\mathcal{M}}\mathcal{M}\right) = 0 \, , \qquad \tilde{\mathcal{M}}\mathcal{M} = \left(\bar{\bar{\eta}}_R\, \gamma^0 M\, \gamma^0 M^+\, \eta_R\right)^T . \tag{3.2}$$

Squared masses of the left- and right-handed (anti)-Weyl fields are given by the eigenvalues of matrices $\mathcal{M}\tilde{\mathcal{M}}$, $\tilde{\mathcal{M}}\mathcal{M}$. Therefore, to generate physical masses the matrices $\mathcal{M}\tilde{\mathcal{M}}$, $\tilde{\mathcal{M}}\mathcal{M}$ must be hermitean and non-negative-definite[3]. We will also require that masses for the left modes coincide with those for the right modes, for only then we can form the standard massive Dirac spinors from their left and right constituents. For the left and the right modes to have the same mass the matrix $\tilde{\mathcal{M}}\mathcal{M}$ must be a similarity transform of $\mathcal{M}\tilde{\mathcal{M}}$ with some non-singular matrix $V$. Therefore, altogether, we require

$$\left(\mathcal{M}\tilde{\mathcal{M}}\right)^+ = \mathcal{M}\tilde{\mathcal{M}},$$
$$\left(\tilde{\mathcal{M}}\mathcal{M}\right)^+ = \tilde{\mathcal{M}}\mathcal{M} \ , \tag{3.3}$$
$$\left(\tilde{\mathcal{M}}\mathcal{M}\right) = V\left(\mathcal{M}\tilde{\mathcal{M}}\right)V^{-1} \ , \qquad \det V \neq 0 .$$

It follows from polar decomposition of arbitrary matrix into a hermitean and unitary factors that the three conditions (3.3) are automatically satisfied when $\tilde{\mathcal{M}} = \mathcal{M}^+$, which is the case for mass matrices in the SM.

In our case, in general, conditions (3.3) are not satisfied. We will now describe the set of all $\mathcal{M}$, $\tilde{\mathcal{M}}$ that satisfy (3.3). As a preliminary step we reduce (3.3) from 4-dimensional to 2-dimensional matrix problem.

Consider equations of motion (3.1). We can use polar decomposition of $\mathcal{M}$ to represent it as a product of two unitary matrices $S_{1,2}$ and a diagonal matrix $M_{diag}$ that has positive entries

$$\mathcal{M} = S_1 \mathcal{M}_{diag} S_2 \, . \tag{3.4}$$

We now decompose each of $S_{1,2}$ into product of two unitary matrices, the first of which, $S_{1+}$, commutes with $\gamma^0 = \Gamma$, i.e., is block-diagonal, while the second one, $S_{1-}$, commutes with $\gamma^0 = \Gamma$ and becomes block-diagonal if we make matrix index swap $1 \leftrightarrow 1,\ 2 \leftrightarrow 3,\ 4 \leftrightarrow 4$

$$S_1 = S_{1+} S_{1-} , \tag{3.5}$$

$$S_2 = S_{2-} S_{2+} , \quad S_{1,2\pm} \in U(2) \oplus U(2), \quad \left[S_{1,2+}, \Gamma\right] = 0 \, , \quad \left[S_{1,2-}, \Gamma\right] \neq 0 \, . \tag{3.6}$$

Such decomposition is always possible: any element of $U(4)$ can be represented in this way. We can now write the equations of motion (3.1) as

[3] Strictly speaking, the eigenvalues can be complex but then they have to have the same phase. We assume that the phase is absorbed in the field redefinition.

$$\left(i\partial\!\!\!/\right)\tilde{\psi}_L - m\left(S_{1-}\mathcal{M}_{diag}\; S_{2-}\right)\tilde{\psi}_R = 0\,, \tag{3.7}$$

$$\left(i\partial\!\!\!/\right)\tilde{\psi}_R - m\left(\Gamma S_{2-}^{+}\mathcal{M}_{diag}\; S_{1-}^{+}\Gamma\right)\tilde{\psi}_L = 0\,, \tag{3.8}$$

where $\tilde{\psi}_L = S_{1+}^{+}\psi_L$, $\tilde{\psi}_R = S_{2+}\psi_R$.

Note now that after index renaming $1 \leftrightarrow 1,\, 2 \leftrightarrow 3,\, 4 \leftrightarrow 4$ matrices $S_{1-}\mathcal{M}_{diag}\; S_{2-}$, $\Gamma S_{2-}^{+}\mathcal{M}_{diag}\; S_{1-}^{+}\Gamma$ become block-diagonal. Hence, we can reduce (3.7-8) to 2-dimensional case for each block of the two matrices. Conditions (3.3) also reduce to two dimensional case with $\Gamma$ replaced by $\Gamma_3 \equiv \sigma_3 = diag(1,-1)$. Written in components an arbitrary $2\times 2$ matrix $\mathcal{M}_R$ and derived from it $\tilde{\mathcal{M}}_R$, where *R* stands for reduced, can be represented as

$$\mathcal{M}_R = \begin{bmatrix} a_{11} & a_{12} \\ a_{21} & a_{22} \end{bmatrix}, \qquad \tilde{\mathcal{M}}_R = \begin{bmatrix} a_{11}^{\;*} & -a_{21}^{\;*} \\ -a_{12}^{\;*} & a_{22}^{\;*} \end{bmatrix}. \tag{3.9}$$

Therefore, the first two conditions in (3.3) may be considered as two linear equations on $a_{21}$, $a_{12}^{*}$

$$\begin{aligned} &a_{11}^{*}\, a_{21} - a_{22}\, a_{12}^{*} = 0\,, \\ &a_{22}^{*}\, a_{21} - a_{11}\, a_{12}^{*} = 0\,. \end{aligned} \tag{3.10}$$

This system has two solutions

$$\begin{aligned} &(1) \quad a_{21} = a_{12} = 0\,,\; a_{11}\,,\, a_{22} \text{ - arbitrary}\,, \\ &(2) \quad a_{21} = \frac{a_{22}}{a_{11}^{*}}\, a_{12}^{*}\,,\; |a_{11}| = |a_{22}|\;. \end{aligned} \tag{3.11}$$

We obtain that the first two conditions (3.3) are satisfied if both $\mathcal{M}\tilde{\mathcal{M}}$ and $\tilde{\mathcal{M}}\mathcal{M}$ are diagonal. In terms of components we obtain

$$\begin{aligned} &\mathcal{M}_R\, \tilde{\mathcal{M}}_R = diag\left(|a_{11}|^2 - |a_{12}|^2, |a_{22}|^2 - |a_{21}|^2\right), \\ &\tilde{\mathcal{M}}_R\, \mathcal{M}_R = diag\left(|a_{11}|^2 - |a_{21}|^2, |a_{22}|^2 - |a_{12}|^2\right). \end{aligned} \tag{3.12}$$

This reduces for solutions (1), (2) to

$$(1)\;\; \mathcal{M}_R\;\tilde{\mathcal{M}}_R = \tilde{\mathcal{M}}_R\;\mathcal{M}_R = diag\left(|a_{11}|^2, |a_{22}|^2\right), \quad a_{11}\,,\, a_{22} \text{ - arbitrary}, \tag{3.13}$$

$$(2)\;\; \mathcal{M}_R\;\tilde{\mathcal{M}}_R = \tilde{\mathcal{M}}_R\;\mathcal{M}_R = 1, \tag{3.14}$$

$$|a_{11}|^2 - |a_{12}|^2 = 1, \quad |a_{11}| = |a_{22}|, \quad |a_{12}| = |a_{21}|, \quad |a_{11}| \ge |a_{12}|\,. \tag{3.15}$$

In both cases the third condition in (3.3) is satisfied automatically. From $\mathcal{M}_R\;\tilde{\mathcal{M}}_R = 1$ for solution (2) we obtain that $\mathcal{M}_R \in U(1,1)$.

In summary, mass matrix $m\mathcal{M}_R$ for solution (1) is diagonal and without loss of generality is given by

$$m\mathcal{M}_R^{(1)} = \begin{pmatrix} \lambda_1 & 0 \\ 0 & -\lambda_2 \end{pmatrix}, \qquad \lambda_1, \lambda_2 \geq 0\,. \tag{3.16}$$

Thus solution (1) splits into two independent solutions with two independent mass parameters. We will call the solution corresponding to mass $m\lambda_1$ a Weyl spinor with scalar spin zero, while the solution corresponding to $-m\lambda_2$ anti-Weyl spinor with scalar spin zero. Taken as a pair the two solutions with $\mathcal{M}_R^{(1)}$ in (3.16) we will be called Dirac-anti-Dirac doublet, or DaD for short, with scalar spin zero. Each of the specific mass solutions corresponds to a single generation out of the four originally contained in the bi-spinor entering (2.10-12). The reasons for such terminology will be explained in Sections 4 and 5.

The second case is more interesting. Since $\mathcal{M}_R$ is determined up to its two Cartan decomposition unitary $U(2)$ factors, the factors can be absorbed in the field redefinition and the simplest way we can represent $\mathcal{M}_R$ for solution (2) is by taking

$$\mathcal{M}_R^{(2)} = \begin{pmatrix} c_\lambda & s_\lambda \\ s_\lambda & c_\lambda \end{pmatrix}, \qquad s_\lambda = \sinh\lambda, \quad c_\lambda = \cosh\lambda\,, \qquad c_\lambda^{\ 2} - s_\lambda^{\ 2} = 1, \quad \lambda \in R\,. \tag{3.17}$$

This matrix is not reducible to a simpler diagonal form, because diagonalizing unitary transformations will not commute with the diagonal $\sigma_3 = diag(1,-1) \in U(1,\,1)$ matrices of the kinetic part of Lagrangian (2.27). Therefore, the plane wave solutions with $\mathcal{M}_R^{(2)}$ are doublets in the generation space. They contain a Weyl spinor and an anti-Weyl spinor that cannot be separated from each other in the sense that free Lagrangian cannot be separated into two parts such that each part contains only one member of the doublet. We will call such solutions as DaD doublet with scalar spin 1/2.

Returning to equations of motion (3.1) we obtain in the 2-dimensional generation space

$$(i\partial\!\!\!/)\tilde{\psi}_L - m\mathcal{M}_R\,\tilde{\psi}_R = 0\,, \tag{3.18}$$

$$(i\partial\!\!\!/)\tilde{\psi}_R - m\hat{\mathcal{M}}_R\,\tilde{\psi}_L = 0\,. \tag{3.19}$$

Here $\hat{\mathcal{M}}_R = \mathcal{M}_R$ is diagonal for the solution (1) and $\hat{\mathcal{M}}_R = \mathcal{M}_R^{-1}$ for solution (2).

We now can completely classify the physically admissible $4\times 4$ mass matrices $\mathcal{M}$ in (3.1). The most general admissible mass matrix is constructed as a direct sum of combinations of two $2\times 2$ diagonal or $U(1,\,1)$ matrices. Accordingly all fermionic mass eigenstates in bi-spinor gauge theory are combinations of DaD pairs with different values of scalar spin. Altogether there are four cases possible. These correspond to distinct mass parameters varying in number from four to two.

In all four cases we can form linear combinations of left and right (anti)-Weyl modes so that their sum forms a doublet of one Dirac and one anti-Dirac spinors. The left/right modes are already in the necessary form for solution (1) when $\hat{\mathcal{M}}_R = \mathcal{M}_R$. The two modes can have different masses. For the case (2) when $\hat{\mathcal{M}}_R = \mathcal{M}_R^{-1}$ we have to redefine the right modes using a $U(1,\ 1)$ transformation: $\tilde{\psi}_R \to \mathcal{M}_R\,\tilde{\psi}_R$ . In the end the two sets of equations of motion for the left/right modes can be combined into a single equation for a DaD doublet $\psi_D^A$

$$\left(i\partial\!\!\!/ - m^A\right)\psi_D{}^A = 0\,, \quad A = 1{,}2\,, \tag{3.20}$$

$$\psi_D{}^A = \left(S_{1+}^+\psi_L\right)^A + \left(\hat{\mathcal{M}}_R\ S_{2+}\psi_R\right)^A, \qquad S_{1,2+} \in U(2)\oplus U(2) \subset U(4), \tag{3.21}$$

where factor $\hat{\mathcal{M}}_R$ appears only for solution (2), where $m^1 = m^2 = m$. Note that although we can formally build up a Dirac spinor from the left and the right Weyl spinors, only for solution (1) the sum (3.21) can be considered as a unitary operation. For solution (2), because then $\hat{\mathcal{M}}_R$ is a non-unitary factor. Therefore, although formally a single Dirac spinor can be constructed, one cannot use it in calculations because the quantum theory with $\psi_D$ from (3.21) is not unitary equivalent to quantum theory with the Weyl pair $\psi_{L,R}$

Altogether, we have four possible cases of mass four-generation matrices $\mathcal{M}$ in (2.27-35) corresponding to the values of scalar spin $(\ p, q\ ),\quad p,q = 0, 1/2$, for DaD pairs and choices of diagonal matrices $\mathcal{M}_R{}^{(1)}$ and matrices $\mathcal{M}_R{}^{(2)} \in U(1{,}1)$. They are given in order of increasing mass degeneracy by

$$1.\ s_s = (\ 0, 0\ ):\quad m\,\mathcal{M} = m_1\,\mathcal{M}_R{}^{(1)} \oplus m_2\,\tilde{\mathcal{M}}_R{}^{(1)}, \tag{3.22}$$

$$2.\ s_s = \left(0, \frac{1}{2}\right):\quad m\,\mathcal{M} = m_1\,\mathcal{M}_R{}^{(1)} \oplus m_2\,\mathcal{M}_R{}^{(2)}, \tag{3.23}$$

$$3.\ s_s = \left(\frac{1}{2}, 0\right):\quad m\,\mathcal{M} = m_2\,\mathcal{M}_R{}^{(2)} \oplus m_1\,\mathcal{M}_R{}^{(1)}, \tag{3.24}$$

$$4.\ s_s = \left(\frac{1}{2}, \frac{1}{2}\right):\quad m\,\mathcal{M} = m_1\,\mathcal{M}_R{}^{(2)} \oplus m_2\,\tilde{\mathcal{M}}_R{}^{(2)}, \tag{3.25}$$

where in all direct sums of matrices the first summand comes with generation indices $A = 1, 3$, while the second with indices $A = 2, 4$. Case 1 describes two DaDs with scalar spin zero, Case 2 one scalar spin zero DaD and one scalar spin ½ DaD, etc. Maximal mass degeneracy can be obtained from case 4 by putting $m_1 = m_2$. In such a case $\mathcal{M} \in U(2{,}2)$. This case was considered in [21].

## 4. **Quantization of the Dirac and the anti-Dirac Spinors.**

Having derived all possible mass terms we can proceed with the canonical quantization procedure. We begin by writing down three different Lagrangians that cover all four possibilities (3.22-25). First action describes a single generation scalar spin zero Dirac spinor

$$S_D = \int d^4x\,\bar{\psi}\,\left(i\partial\!\!\!/ - m\right)\psi\,. \tag{4.1}$$

Second action describes a single generation scalar spin zero anti-Dirac spinor

$$S_{aD} = -\int d^4x\,\bar{\psi}\,\left(i\partial\!\!\!/ - m\right)\psi\,. \tag{4.2}$$

Third action describes a generation doublet scalar spin $1/2$ DaD spinor

$$S_{DaD}=\int d^4x\ \bar{\psi}^A\Gamma_3{}^{AB}\left(i\partial\!\!\!/-m\mathcal{M}\right)\psi^B\ ,\ \ A,B=1,2\,,\qquad \Gamma_3\equiv\sigma_3=\mathrm{diag}\left(\,1,-1\right), \tag{4.3}$$

where $\mathcal{M}\in U\left(1,\,1\right)$ is given by (3.17) and where, in order to avoid confusion between $2\times 2$ matrices acting on the Lorentz indices and on the generation indices, we renamed Pauli matrices acting on the generation indices as $\Gamma_k\equiv\sigma_k$, $\Gamma_k\Gamma_l=\delta_{kl}-i\varepsilon_{klm}\Gamma_m$. The third action cannot be diagonalized further, because the unitary transformation needed to diagonalize the mass term in (4.3) does not commute with $\Gamma_3$. This means that we cannot consider field components $A,B=1,2$ as representing independent free fields. Instead, we have to treat $\psi^A$ as a two component field describing a single particle with an additional degree of freedom, described by an additional quantum number, which we will refer to as scalar spin quantum number. Explanation for the name will be clear from the discussion below.

We will now proceed with the canonical quantization procedure applied to the three Lagrangians in (4.1-3). A preparatory step for canonical quantization is the expressing of on-shell fields in terms of superpositions of plane-wave solutions of free field equations of motion in such a way that the corresponding Hamiltonians are positive definite diagonal bilinears in the amplitudes of plane wave expansions. This is followed by interpretation of the amplitudes of the expansion as creation and annihilation operators acting in the Fock space, which is constructed as space of polynomials of creation operators acting on the vacuum state, which is defined as that annihilated by all annihilation operators.

For the standard Dirac spinor action (4.1) this preliminary step is also standard. Nevertheless, we will write out the procedure explicitly to set notations and to have background expressions against which we can compare non-trivial differences of quantization between (4.1) and (4.2-3). For scalar spin zero anti-Dirac spinor in (4.2) plane wave expansion will be also standard. However, the assignment of the creation and annihilation operators will have to be swapped to ensure positivity of the Hamiltonian. The amplitudes in plane wave expansion of free fields in (4.3) will be definitely non-standard, since in addition to spin quantum number they will also carry the scalar spin quantum number.

We begin with the standard expansion for (4.1) which is given by

$$\psi_D\left(x\right)=\int\frac{d^3k}{\left(2\pi\right)^3}\frac{m}{k^0}\left(b_r\left(\vec{k}\right)\ u^r\left(\vec{k}\right)e^{-ikx}+d_r{}^{+}\left(\vec{k}\right)v^r\left(\vec{k}\right)e^{ikx}\right), \tag{4.4}$$

$$\bar{\psi}_D\left(x\right)=\int\frac{d^3k}{\left(2\pi\right)^3}\frac{m}{k^0}\left(b_r{}^{+}\left(\vec{k}\right)\bar{u}^r\left(\vec{k}\right)e^{ikx}+d_r\left(\vec{k}\right)\bar{v}^r\left(\vec{k}\right)e^{-ikx}\right), \tag{4.5}$$

where $k^0=\sqrt{\vec{k}^2+m^2}$ and were we follow the normalization conventions of [24]. The plane-wave solutions $u^r\left(\vec{k}\right),\ r=1,2$, for positive and $v^r\left(\vec{k}\right),\ r=1,2$, for negative energy satisfy phase space Dirac equations of motion $\left(k\!\!\!/-m\right)u^r\left(\vec{k}\right)=\left(k\!\!\!/+m\right)v^r\left(\vec{k}\right)=0$ and are normalized according to

$$\begin{aligned}&\bar{u}^r\left(\vec{k}\right)u^s\left(\vec{k}\right)=\delta^{rs}\,,\qquad \bar{v}^r\left(\vec{k}\right)v^s\left(\vec{k}\right)=-\delta^{rs}\,,\qquad \bar{u}^r\left(\vec{k}\right)v^s\left(\vec{k}\right)=0\,,\\ &u^r{}_\alpha\left(\vec{k}\right)\bar{u}^r{}_\beta\left(\vec{k}\right)=\frac{\left(k\!\!\!/+m\right)_{\alpha\beta}}{2m}\,,\qquad v^r{}_\alpha\left(\vec{k}\right)\bar{v}^r{}_\beta\left(\vec{k}\right)=\frac{\left(k\!\!\!/-m\right)_{\alpha\beta}}{2m}\,.\end{aligned} \tag{4.6}$$

They are chosen to form states with definite helicity projections on $\vec{k}$ : $u^1(\vec{k})$ and $v^2(\vec{k})$ correspond to helicity $1/2$ , while $u^2(\vec{k})$ and $v^1(\vec{k})$ correspond to states with helicity $-1/2$ .

Canonical quantization replaces classical Grassmann-valued amplitudes in (4.4-5) with quantum operators in Fock space with the canonical anticommutation relations

$$\left\{b_r(\vec{k}), b_s^{+}(\vec{k}')\right\} = (2\pi)^3 \frac{k^0}{m} \delta_{rs} \delta^3(\vec{k} - \vec{k}'),$$
$$\left\{d_r(\vec{k}), d_s^{+}(\vec{k}')\right\} = (2\pi)^3 \frac{k^0}{m} \delta_{rs} \delta^3(\vec{k} - \vec{k}'), \tag{4.7}$$

while the remaining anticommutators remain zero. In (4.7) operators $b_r^{+}(\vec{k}),\ b_s(\vec{k})$ are creation and annihilation operators for Dirac particles, while $d_r^{+}(\vec{k}),\ d_s(\vec{k})$ are creation and annihilation operators for Dirac antiparticles. All act on Fock space which consists of polynomials of creation operators $P(b_r^{+}(\vec{k}), d_r^{+}(\vec{k}))$ acting on the vacuum state $|0\rangle$, defined as the state that is annihilated by all annihilation operators. From (4.1, 4.4-6) we obtain total Dirac energy momentum operator $P^{\mu}{}_D$ and Dirac $U(1)$ charge operator $Q_D$

$$P^{\mu}{}_D =: \int \frac{d^3k}{(2\pi)^3} \frac{m}{k^0} k^{\mu} \left(b_r^{+}(\vec{k})\, b_r(\vec{k}) + d_r^{+}(\vec{k})\, d_r(\vec{k})\right):, \qquad \langle 0|P^{\mu}{}_D|0\rangle \geq 0\,, \tag{4.8}$$

$$Q_D =: \int \frac{d^3k}{(2\pi)^3} \frac{m}{k^0} \left(b_r^{+}(\vec{k}) b_r(\vec{k}) - d_r^{+}(\vec{k})\, d_r(\vec{k})\right):, \tag{4.9}$$

where : : denotes the normal ordering of the operators. The time-ordered product of two Dirac fields, defined by

$$T\left[\psi_\alpha(x)\bar{\psi}_\beta(y)\right] = \theta(x^0 - y^0)\psi_\alpha(x)\bar{\psi}_\beta(y) - \theta(y^0 - x^0)\bar{\psi}_\beta(y)\psi_\alpha(x), \tag{4.10}$$

after separation into its vacuum expectation value and the normal-ordered parts according to

$$T\left[\psi(x)\bar{\psi}(y)\right] = \langle 0|\psi(x)\bar{\psi}(y)|0\rangle + :\psi(x)\bar{\psi}(y):, \tag{4.11}$$

defines the Feynman propagator $S_F(z)$ for Dirac field

$$S_F(x-y) = \langle 0|\psi(x)\bar{\psi}(y)|0\rangle = \int \frac{d^4k}{(2\pi)^4} S_F(k) e^{-ik(x-y)},$$
$$S_F(k) = \frac{i(\not{k} + m)}{k^2 - m^2 + i\varepsilon}, \tag{4.12}$$

where $+i\varepsilon$ in the denominator specifies poles of $S_F(k)$ in the complex $k^0$ plane. This concludes an outline of the canonical quantization of Dirac field.

Following the same steps we will now describe quantization of anti-Dirac field with action $S_{aD}$ given by (4.2). Since $S_{aD} = -S_D$, we obtain for classical fields

$$P^{\mu}{}_{aD} = -P^{\mu}{}_{D},$$
$$Q_{aD} = -Q_D. \tag{4.13}$$

Obviously, we can use the same plane-wave expansion (4.4-6) as in the Dirac case but we cannot use the Dirac field quantization (4.7). The interpretation of $b_r^{+}(\vec{k}),\ b_s(\vec{k})$ as creation and annihilation operators for Dirac particles and $d_r^{+}(\vec{k}),\ d_s(\vec{k})$ as creation and annihilation operators for Dirac anti-particles results in non-positive Hamiltonian $H_{aD} = P^0{}_{aD}$. This fact was the cause for the long held belief that bi-spinor theory is consistent only in Euclidean space-times. As we will now see, this belief was unfounded.

It is easy to cure the problem. The hint comes from $Q_{aD} = -Q_D$, which suggests that for anti-Dirac spinors particles and antiparticles should be swapped. Since classical Dirac spinor Hamiltonian is indefinite in any case, all we have to do is to redefine the notions of what are the amplitudes for particle and anti-particle and instead of (4.4-5) use a modified plane-wave expansion

$$\psi_{aD}(x) = \int \frac{d^3k}{(2\pi)^3}\frac{m}{k^0}\left(b_r^{+}(\vec{k})u^r(\vec{k})\,e^{-ikx} + d_r(\vec{k})v^r(\vec{k})\,e^{ikx}\right), \tag{4.14}$$

$$\bar{\psi}_{aD}(x) = \int \frac{d^3k}{(2\pi)^3}\frac{m}{k^0}\left(b_r(\vec{k})\bar{u}^r(\vec{k})\,e^{ikx} + d_r^{+}(\vec{k})\bar{v}^r(\vec{k})\,e^{-ikx}\right), \tag{4.15}$$

where now $u^1(\vec{k})$ and $v^2(\vec{k})$ correspond to states with helicity $-1/2$ , while $u^2(\vec{k})$ and $v^1(\vec{k})$ correspond to states with helicity $1/2$.

After the change the expressions (4.6-9) remain the same and we obtain the desired result for quantum anti-Dirac field. The energy momentum and the charge operators have the same form as the anti-Dirac creation-annihilation operator assignment

$$P^{\mu}{}_{aD}\left(b_r, b_r^{+}, d_r, d_r^{+}\right) = P^{\mu}{}_{D}\left(b_r^{+}, b_r, d^{+}{}_r, d_r\right), \qquad \langle 0|P^{\mu}{}_{aD}|0\rangle \geq 0,$$
$$Q_{aD}\left(b_r, b_r^{+}, d_r, d_r^{+}\right) = Q_D\left(b_r^{+}, b_r, d^{+}{}_r, d_r\right). \tag{4.16}$$

The minus in the action (4.2) and subsequent reassignment of creation and annihilation operators for anti-Dirac spinors leads to potentially significant physical consequences, because it induces a notable difference between the Dirac Feynman and the anti-Dirac Feynman propagators. One obtains what we will call anti-Feynman phase space propagator

$$S_{aF}(k) = \frac{i}{-(\not{k} - m) + i\varepsilon} = -i\frac{(\not{k} + m)}{k^2 - m^2 - i\varepsilon}, \tag{4.17}$$

which should be compared with the expression for the Dirac case where

$$S_F(k)=\frac{i}{+(\not k-m)+i\varepsilon}=i\frac{(\not k+m)}{k^2-m^2+i\varepsilon}. \tag{4.18}$$

Note that the sign of $+i\varepsilon$ in (4.17-18) is fixed by the requirement that the large time limit of Schrödinger evolution of free-field vacuum state with full Hamiltonian results in the ground state of the full Hamiltonian. This, in turn, is a necessary condition for the existence of the perturbation theory. We see that, apart from phase factor $(-1)$ the difference between the two propagators is in the position of their poles. Positions of poles of $S_{aF}(k)$ and $S_F(k)$ are shown on Fig.1.

Functions $S_{aF}(k)$, $S_F(k)$ are related, for the sum $S_F(k)+S_{aF}(k)$ can be computed using the Cauchy residue theorem or Sokhotski formula about distributions

$$\lim_{\varepsilon\to 0^+}\frac{1}{x\pm i\varepsilon}=P\left(\frac{1}{x}\right)\mp i\pi\delta(x), \tag{4.19}$$

where $P$ denotes the Cauchy principle value of the integral, also called the principle part of the integral. We obtain

$$K(k)\equiv S_F(k)+S_{aF}(k)=2\pi(\not k+m)\delta(k^2-m^2)=\int d^4x e^{+ikx}(i\not\partial+m)J(x),$$

(4.20)

$$J(x)=\int\frac{d^4k}{(2\pi)^3}\delta(k^2-m^2)\,e^{-ikx}=\int\frac{d^3k}{(2\pi)^3}\frac{\cos(E_k x^0)}{E_k}e^{i\vec{k}\vec{x}},\qquad E_k=\sqrt{\vec{k}^2+m^2}.$$

The integral in (4.20) can be evaluated in quadratures for the case when $m=0$. With the help of

$$\int_0^\infty dk\frac{\sin(x^0 k)\sin(|\vec{x}|k)}{k}=\frac{1}{4}\operatorname{sign}(x^0)\ln\left(\frac{|x^0|+|\vec{x}|}{|x^0|-|\vec{x}|}\right)^2, \tag{4.21}$$

we obtain an expression that is relativistically invariant with regard to proper Lorentz group but is not time-reversal invariant

$$J(x)=-\frac{1}{2\pi^2}\frac{1}{x^2}\operatorname{sign}(x^0),\qquad x^2=x^\mu x_\mu. \tag{4.22}$$

From (4.22) we obtain for the massless case an explicit expression

$$K(x)=S_F(x)+S_{aF}(x)=-\frac{i}{2\pi^2}\not\partial\left(\frac{1}{x^2}\operatorname{sign}(x^0)\right). \tag{4.23}$$

The massive case can also be calculated explicitly. Using

$$\int_0^\infty dk\frac{\cos\left(x^0\sqrt{k^2+m^2}\right)\cos(|\vec{x}|k)}{\sqrt{k^2+m^2}}=-\frac{\pi}{2}\theta(x^2)Y_0\left(m(x^2)^{1/2}\right)+\theta(-x^2)K_0\left(m(-x^2)^{1/2}\right), \tag{4.24}$$

where $\theta(z)=0 \quad z<0$, $\theta(z)=1 \quad z\geq 0$ is the step function, and $Y_0(z)$, $K_0(z)$ are the 0-th Neumann function and 0-th Bessel function of imaginary argument, respectively, we obtain

$$K(x)=S_F(x)+S_{aF}(x)$$
$$=\frac{m}{2\pi^2}(i\partial\!\!\!/+m)\left\{\frac{1}{|x^2|^{1/2}}\left(\frac{\pi}{2}\theta(x^2)Y_0'\left(m(x^2)^{1/2}\right)+\theta(-x^2)K_0'\left(m(-x^2)^{1/2}\right)\right)\right\}, \tag{4.25}$$

where $Y_0'(z)$, $K_0'(z)$ are derivatives with respect to the argument.

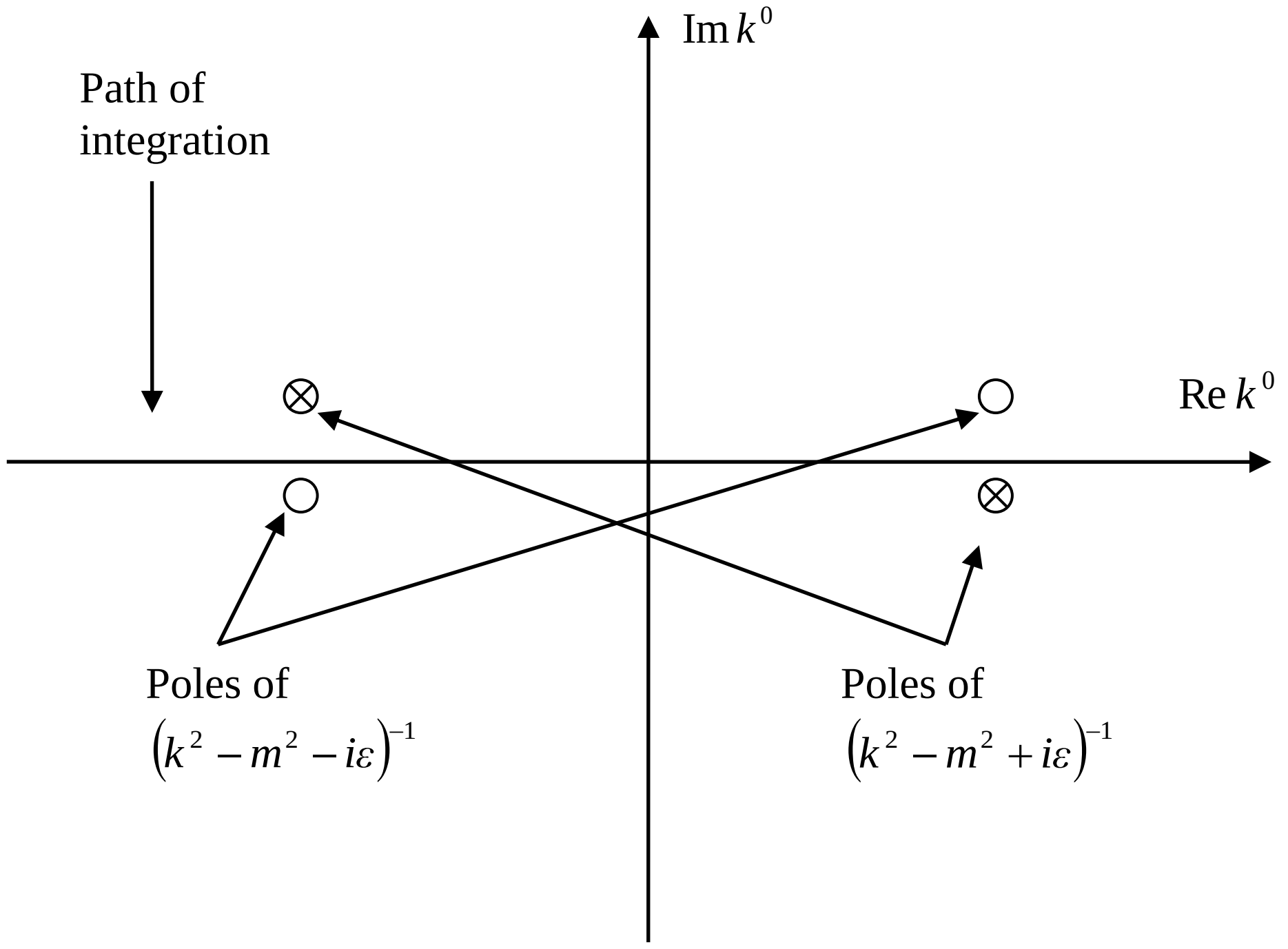

Fig. 1. Poles of Dirac and anti-Dirac propagators.

Since in principle there could be a physical difference between Dirac and anti-Dirac particles in bi-spinor gauge theory, in bi-spinor gauge theories we have to use four distinct fermionic particle types: Dirac particles, Dirac anti-particles, anti-Dirac particles, and anti-Dirac anti-particles. From $(k\!\!\!/-m)K(k)=2\pi(k^2-m^2)\delta(k^2-m^2)=0$ we obtain that $K(x)$is a solution of Dirac equation: $(i\partial\!\!\!/-m)K(x)=0$, that is $S_F(k)$, $S_{aF}(k)$ differ by a phase factor $(-1)$ up to a solution of Dirac equation. Although this is expected, the fact has an important consequence. Intuitively, the difference should not make S-matrix elements computed between in and out states for anti-Dirac particles different from those for Dirac particles. After all, all we did to introduce anti-Dirac fermions was to exchange particles with anti-particles. We will now show that this is not the case, if anti-Dirac particles couple to integer spin fields or Dirac particles. However, the difference between the scattering amplitudes comes from the terms that appear only in loop momentum integrals.

Note that the appearance of $K(x)$ does not play role in external lines. To see this let us consider the LSZ reduction formula and the structure of perturbation expansion of time ordered products of interacting fields in terms of propagators and vertex functions. Consider two states in Fock space. The in-state has $n_i$ fermionic particles with momenta labeled $(k_1,\dots,k_{n_i})$ and $m_i$ antiparticles with momenta labeled $(k'_1,\dots,k'_{m_i})$. The out-state has $n_o$ fermionic particles with momenta labeled $(p_1,\dots,p_{n_o})$ and $m_o$ antiparticles with momenta labeled $(p'_1,\dots,p'_{m_o})$. The transition amplitude between the two fermionic states is given by the LSZ reduction formula

$$
\begin{aligned}
&\langle out|\, d_{out}(p'_1)\cdots d_{out}(p'_{m_o})\, b_{out}(p_1)\cdots b_{out}(p_{n_o})\, b^{+}{}_{in}(k_1)\cdots b^{+}{}_{in}(k_{n_i})\, d^{+}{}_{in}(k'_1)\cdots d^{+}{}_{in}(k'_{n_i})\,|in\rangle \\
&= (-i)^{n_i+n_o}(+i)^{m_i+m_o}\int d^4x_1\cdots d^4x_{n_i}\, d^4y_1\cdots d^4y_{m_i}\, d^4x'_1\cdots d^4x'_{n_o}\, d^4y'_1\cdots d^4y'_{m_o} \\
&\times \exp\left(-ik_l x_l - k'_l\, x'_l + ip_l y_l + p'_l\, y'_l\right) \\
&\times \bar{u}(p_1)\left(i\vec{\partial}_{y_1}-m\right)\cdots \bar{u}(p_{n_i})\left(i\vec{\partial}_{y_{n_i}}-m\right)\bar{v}(k'_1)\left(i\vec{\partial}_{x'_1}-m\right)\cdots \bar{v}(k'_{m_i})\left(i\vec{\partial}_{x'_{m_i}}-m\right) \qquad (4.26) \\
&\times \langle 0|T\left[\overline{\psi}(y'_{m_o})\cdots\overline{\psi}(y'_1)\psi(y_{n_o})\cdots\psi(y_1)\overline{\psi}(x_1)\cdots\overline{\psi}(x_{n_i})\psi(x'_1)\cdots\psi(x'_{m_i})\right]|0\rangle \\
&\times \left(-i\overleftarrow{\partial}_{x_1}-m\right)\bar{u}(p_1)\cdots\left(-i\overleftarrow{\partial}_{x_{n_o}}-m\right)\bar{u}(p_{n_i})\left(-i\overleftarrow{\partial}_{y'_o}-m\right)\bar{v}(p'_1)\cdots\left(-i\overleftarrow{\partial}_{y'_{m_o}}-m\right)\bar{v}(p'_{m_o}) \\
&+ disconnected\ \ part.
\end{aligned}
$$

We need to consider only the connected part, because the disconnected part is a sum of products of connected parts to each of which our argument applies.

## 5. Quantization of the DaD Spinors

We will now turn to the canonical quantization of DaD particles with scalar spin 1/2, where changes in quantization are more prominent. We begin with writing the free Lagrangian for the DaD doublet. The free field DaD Lagrangian density (4.3) is

$$
\mathcal{L}_{DaD} = \overline{\overline{\psi}}_L (i\partial)\psi_L + \overline{\overline{\psi}}_R (i\partial)\psi_R - m\left(\overline{\overline{\psi}}_L \mathcal{M}\psi_R + \overline{\overline{\psi}}_R \tilde{\mathcal{M}}\,\psi_L\right). \qquad (5.1)
$$

The Cartan decomposition of a $U(1,1)$ matrix $\mathcal{M}$ is given by

$$
\begin{aligned}
&\mathcal{M} = U_1 R U_2\,,\quad U_{1,2} = diag\left(\exp(i\alpha_{1,2}),\ \exp(i\beta_{1,2})\right), \\
&R = \begin{pmatrix} c_\lambda & s_\lambda \\ s_\lambda & c_\lambda \end{pmatrix},\quad c_\lambda = \cosh\lambda,\quad s_\lambda = \sinh\lambda\,.
\end{aligned} \qquad (5.2)
$$

The choice of $U_{1,2}$ in the decomposition is unique up to multiplication by a phase. We now redefine $\psi_{L,R}$ to absorb $U_{1,2}$: $\tilde{\psi}_L = U_1^{+} P_L \psi$, $\tilde{\psi}_R = U_2 P_R \psi$ to obtain in terms of Dirac spinors $\psi^A = \left(\psi_L{}^A + \psi_R{}^A\right)$, $A = 1,2$,

$$\mathcal{L}_{DaD} = \overline{\psi}^1\left(i\partial\!\!\!/ - m\,c_\lambda\right)\psi^1 - \overline{\psi}^2\left(i\partial\!\!\!/ - m\,c_\lambda\right)\psi^2 - m\,s_\lambda\ \varepsilon^{AB}\overline{\psi}^A\gamma^5\psi^B, \tag{5.3}$$

We will now bring (5.3) into a more convenient form by a $O(2)$ rotation of $\psi^A$. We define new fields $\xi^A$ by

$$\begin{pmatrix}\psi^1\\ \psi^2\end{pmatrix} = \frac{1}{\sqrt{2}}\begin{pmatrix}1 & 1\\ -1 & 1\end{pmatrix}\begin{pmatrix}\xi^1\\ \xi^2\end{pmatrix}. \tag{5.4}$$

In terms of the new fields we obtain

$$\mathcal{L}_{DaD} = \left(\overline{\xi}^1, \overline{\xi}^2\right)\begin{pmatrix}0 & \left(i\partial\!\!\!/ - mc_\lambda - ms_\lambda\ \gamma^5\right)\\ \left(i\partial\!\!\!/ - mc_\lambda + ms_\lambda\ \gamma^5\right) & 0\end{pmatrix}\begin{pmatrix}\xi^1\\ \xi^2\end{pmatrix}. \tag{5.5}$$

The equations of motion for the new fields decouple

$$\begin{aligned}&\left(i\partial\!\!\!/ - m(c_\lambda + s_\lambda)\,\gamma^5\right)\xi^1 = 0\,,\\ &\left(i\partial\!\!\!/ - m(c_\lambda - s_\lambda)\,\gamma^5\right)\xi^2 = 0\,.\end{aligned} \tag{5.6}$$

Note that we only have to solve the first equation. Solutions of the second are obtained by replacing $s_\lambda \to -s_\lambda$. Recalling that the mass scale parameters $s_\lambda = 1/2\left(e^\lambda - e^{-\lambda}\right)$, $c_\lambda = 1/2\left(e^\lambda + e^{-\lambda}\right)$, we see that the $\xi^A$ doublet contains a free massive Dirac spinor with left/right masses $e^\lambda, -e^\lambda$ for $\xi^1$ and a free massive Dirac spinor with left/right masses $e^{-\lambda}, -e^{-\lambda}$ for $\xi^2$. In fact the doublet is a sum of the scalar spin $1/2$ and $-1/2$ states.

The detailed plane wave solutions of the decoupled equations are described in Appendix A. In Appendix B we derive the correct basis for creation and annihilation operators. They provide for the normalized phase space basis $u^A_{rp}\left(\vec{k}\right)$, $v^A_{rp}\left(\vec{k}\right)$ of the expansion of the quantized fields. The plane wave expansion of the DaD doublet $\psi^A(x)$, $A = 1, 2$, and the non-zero canonical anti-commutation relations are given by

$$\begin{aligned}\psi^A(x) &= \int \frac{d^3k}{(2\pi)^3}\frac{m}{k^0}\sum_{s,p=1,2}\left[u^A_{rp}\left(\vec{k}\right)b_{rp}\left(^+\right)^{p-1}\left(\vec{k}\right)e^{-ikx} + v^A_{rp}\left(\vec{k}\right)d_{rp}\left(^+\right)^{p}\left(\vec{k}\right)e^{ikx}\right],\\ \overline{\psi}^A(x) &= \int \frac{d^3k}{(2\pi)^3}\frac{m}{k^0}\sum_{s,p=1,2}\left[\overline{u}^A_{rp}\left(\vec{k}\right)b_{rp}\left(^+\right)^{p}\left(\vec{k}\right)e^{+ikx} + \overline{v}^A_{rp}\left(\vec{k}\right)d_{rp}\left(^+\right)^{p-1}\left(\vec{k}\right)e^{-ikx}\right],\end{aligned} \tag{5.7}$$

$$\begin{aligned}\left\{b_{sp}\left(\vec{k}\right), b_{qr}{}^+\left(\vec{k}'\right)\right\} &= (2\pi)^3\frac{k^0}{m}\delta_{sq}\delta_{pr}\delta^3\left(\vec{k} - \vec{k}'\right),\\ \left\{d_{sp}\left(\vec{k}\right), d_{qr}{}^+\left(\vec{k}'\right)\right\} &= (2\pi)^3\frac{k^0}{m}\delta_{sq}\delta_{pr}\delta^3\left(\vec{k} - \vec{k}'\right),\end{aligned} \tag{5.8}$$

where notation $\left(^{+}\right)^{p}$ means Hermitean conjugation in power $p$ : $b\left(^{+}\right)^{1}\equiv b^{+}, b\left(^{+}\right)^{2}\equiv\left(b^{+}\right)^{+}=b$, etc. Note that the creation/annihilation operators in (5.7, 5.8) acquired an additional index for the value of scalar spin quantum number. In order to avoid confusion when dealing with DaD creation/annihilation operators, the first index of the operator will always describe Lorentz spin states, while the second the scalar spin states.

Plane-wave solutions $u^{A}_{r\,p}\left(\vec{k}\right)$, which form the complete orthonormal set of solutions of free field equations of motion, depend on the generation index $A$, the spin index $r$, and on the scalar spin index $p$, each of which takes two values. They are obtained with the help of two unitary transformations. The first unitary transformation, $T_x=\left\{T_x{}^{A}{}_{B}\right\},\ A,B=1,2,$ is a constant rotation of two generations in the generation space, while the second unitary transformation $T_k=\left\{T_k{}^{p}{}_{q}\right\}$ $p,q=1,2,$ is a constant rotation in the two dimensional scalar spin space. The coefficients of the two transformations are related by the consistency condition that requires that in the case of $\mathcal{M}=1$ the combination of the two transformations is identity, i.e., $T_xT_k=1$, from which we obtain that $T_k=T_x^{-1}$. When $\mathcal{M}\neq 1$ the plane wave solutions are given by

$$u^{A}_{r\,p}\left(\vec{k}\right)=-\left(T_x\right)^{A}_{B}\,\xi^{(+)B}_{rq}\left(\vec{k},s_\lambda\right)\left(T_k\right)^{q}{}_{p}\,,$$
$$v^{A}_{r\,p}\left(\vec{k}\right)=-\left(T_x\right)^{A}_{B}\,\xi^{(-)B}_{rq}\left(\vec{k},s_\lambda\right)\left(T_k\right)^{q}{}_{p}\,, \tag{5.9}$$

where mass scale parameter $s_\lambda\equiv\sinh\lambda$ is defined in (3.17) and $\xi^{(\pm)B}_{rq}\left(\vec{k},s_\lambda\right)$ are given by

$$\hat{\xi}^{(\pm)}{}_{r1}\left(\vec{k}\right)=\begin{pmatrix}\xi^{(\pm)}_{r}\left(\vec{k},s_\lambda\right)\\ 0\end{pmatrix},\qquad \hat{\xi}^{(\pm)}{}_{r2}\left(\vec{k}\right)=\begin{pmatrix}0\\ \xi^{(\pm)}_{r}\left(\vec{k},-s_\lambda\right)\end{pmatrix}. \tag{5.10}$$

The details of the derivation are presented in Appendix A. Matrices $T_x$, $T_k$ and the normalized basis for the configuration space fields in (5.7) are given by

$$T_x=\frac{1}{\sqrt{2}}\begin{pmatrix}1&1\\-1&1\end{pmatrix},\qquad T_k=\frac{1}{\sqrt{2}}\begin{pmatrix}1&-1\\1&1\end{pmatrix}=T_x^{-1}, \tag{5.11}$$

$$\begin{pmatrix}u^{1}_{r1}\left(\vec{k}\right)\\ u^{2}_{r1}\left(\vec{k}\right)\end{pmatrix}=\frac{1}{2}\begin{pmatrix}+\xi^{(+)}{}_{r}\left(\vec{k},s_\lambda\right)-\xi^{(+)}{}_{r}\left(\vec{k},-s_\lambda\right)\\ -\xi^{(+)}{}_{r}\left(\vec{k},s_\lambda\right)-\xi^{(+)}{}_{r}\left(\vec{k},-s_\lambda\right)\end{pmatrix},\quad \begin{pmatrix}u^{1}_{r2}\left(\vec{k}\right)\\ u^{2}_{r2}\left(\vec{k}\right)\end{pmatrix}=\frac{1}{2}\begin{pmatrix}+\xi^{(+)}{}_{r}\left(\vec{k},s_\lambda\right)+\xi^{(+)}{}_{r}\left(\vec{k},-s_\lambda\right)\\ -\xi^{(+)}{}_{r}\left(\vec{k},s_\lambda\right)+\xi^{(+)}{}_{r}\left(\vec{k},-s_\lambda\right)\end{pmatrix},$$
$$\begin{pmatrix}v^{1}_{r1}\left(\vec{k}\right)\\ v^{2}_{r1}\left(\vec{k}\right)\end{pmatrix}=\frac{1}{2}\begin{pmatrix}+\xi^{(-)}{}_{r}\left(\vec{k},s_\lambda\right)-\xi^{(-)}{}_{r}\left(\vec{k},-s_\lambda\right)\\ -\xi^{(-)}{}_{r}\left(\vec{k},s_\lambda\right)-\xi^{(-)}{}_{r}\left(\vec{k},-s_\lambda\right)\end{pmatrix},\quad \begin{pmatrix}v^{1}_{r2}\left(\vec{k}\right)\\ v^{2}_{r2}\left(\vec{k}\right)\end{pmatrix}=\frac{1}{2}\begin{pmatrix}+\xi^{(-)}{}_{r}\left(\vec{k},s_\lambda\right)+\xi^{(-)}{}_{r}\left(\vec{k},-s_\lambda\right)\\ -\xi^{(-)}{}_{r}\left(\vec{k},s_\lambda\right)+\xi^{(-)}{}_{r}\left(\vec{k},-s_\lambda\right)\end{pmatrix}. \tag{5.12}$$

When $\mathcal{M}$ is diagonal then $s_\lambda=0$ and, as expected, in (5.8) $T_x, T_k$ cancel each other. We then obtain a doublet of two decoupled Dirac and anti-Dirac spinors, each a state with scalar spin zero. The normalized 8-component DaD positive/negative energy solutions of DaD equations of motions in (5.12) satisfy

$$\bar{u}^A_{rp}(\vec{k})\, u^A_{r'q}(\vec{k}) = \delta_{rr'}\delta_{pq},$$

$$\bar{v}^A_{rp}(\vec{k})\, v^A_{r'q}(\vec{k}) = -\delta_{rr'}\delta_{pq},$$

$$\bar{u}^A_{rp}(\vec{k})\, v^A_{r'q}(\vec{k}) = 0, \tag{5.13}$$

$$\left(\bar{u}^A_{rp}(\vec{k})\right)_\alpha \left(u^B_{rp}(\vec{k})\right)_\beta = \frac{1}{2m}\left(\left((\not{k} + mc_\lambda)\Gamma_3^{AB} + ims_\lambda \Gamma_1^{AB}\gamma^5\right)\right)_{\alpha\beta},$$

$$\left(\bar{v}^A_{rp}(\vec{k})\right)_\alpha \left(v^B_{rp}(\vec{k})\right)_\beta = \frac{1}{2m}\left(\left((\not{k} - mc_\lambda)\Gamma_3^{AB} + ims_\lambda \Gamma_1^{AB}\gamma^5\right)\right)_{\alpha\beta}.$$

For mass parameters $s_\lambda = 0,\ c_\lambda = 1$ we recover the standard expressions for one Dirac and one anti-Dirac spinors. The details of derivation are in Appendix A.

Using plane-wave expansion (5.7) we obtain diagonal bilinear forms for the energy-momentum and the $U(1)$ charge of scalar spin 1/2 DaD field

$$P^\mu = :\int \frac{d^3k}{(2\pi)^3}\frac{m}{k^0} k^\mu \left(b_{sp}{}^+(\vec{k})\, b_{sp}(\vec{k}) + d_{sp}{}^+(\vec{k})\, d_{sp}(\vec{k})\right):, \tag{5.14}$$

$$Q = :\int \frac{d^3k}{(2\pi)^3}\frac{m}{k^0} \left(b_{sp}{}^+(\vec{k})\, b_{sp}(\vec{k}) - d_{sp}{}^+(\vec{k})\, d_{sp}(\vec{k})\right):. \tag{5.15}$$

The physical meaning of the scalar spin index of the creation/annihilation operators is discussed in detail in Appendix C.

The last ingredient for perturbation theory with DaD bi-spinors is an expression for its propagator. Following the standard prescription for extraction of the propagator from the bilinear part of the free action we obtain what we will call the DaD propagator

$$S_\lambda{}^{AB}(k) = i\left((\not{k} - mc_\lambda)\Gamma_3^{AB} - i\, ms_\lambda \Gamma_1^{AB}\gamma^5 + i\varepsilon\delta^{AB}\right)^{-1}. \tag{5.16}$$

When $s_\lambda = 0,\ c_\lambda = 1$, as expected, propagator (5.16) becomes a propagator for a scalar spin zero doublet of one generation of Dirac and one generation of anti-Dirac spinors

$$S_{\lambda=0}{}^{AB}(k) = i\left((\not{k} - m)\Gamma_3^{AB} + i\varepsilon\delta^{AB}\right)^{-1} = \begin{pmatrix} S_F(k) & 0 \\ 0 & S_{aF}(k) \end{pmatrix}. \tag{5.17}$$

Since we have expressions for propagators and the Fock space for the Dirac and anti-Dirac (anti)particles is constructed according to the standard rules of quantum field theory, to construct formal perturbation theory we only need to write down Feynman rules for vertices. The remaining Feynman rules for gauge fields and ghosts remain unaffected by the presence of anti-Dirac particles.

Therefore, apart from the propagators the only new ingredients bi-spinors bring in are the two new vertex functions: one for the anti-Dirac spinor and one for the DaD doublet. The two vertices can be read off the interaction Lagrangians obtained from free field expression (5.3) by minimal gauging procedure. The two Lagrangians are given by

$$\mathcal{L}_{aD} = -\overline{\psi}_{aD}\left(i\partial\!\!\!/ - m + gA\!\!\!/\right)\psi_{aD}, \tag{5.18}$$

$$\mathcal{L}_{DaD} = \overline{\psi}_{DaD}\Gamma_3\left(i\partial\!\!\!/ - m\mathcal{M} + gA\!\!\!/\right)\psi_{DaD}, \tag{5.19}$$

where $g$ is the coupling constant and $A\!\!\!/ = \gamma^{\mu} A_{\mu}{}^{a}\tau^{a}$ is the $\gamma$-matrix contracted form of gauge field $A_{\mu}$ with Lie algebra generators $\tau^{a}$. From (5.18) we obtain that the vertex function for anti-Dirac case with one $\psi_{aD}$ incoming and one $\overline{\psi}_{aD}$ outgoing is obtained from that of Dirac case by changing the sign of the coupling constant $g$. Therefore, we obtain for the Dirac and the anti-Dirac cases

$$\mathcal{V}_{D} = g\gamma^{\mu}\tau^{a}, \tag{5.20}$$

$$\mathcal{V}_{aD} = -g\gamma^{\mu}\tau^{a}. \tag{5.21}$$

Vertex function for the DaD case is a bit more complicated. We obtain

$$\mathcal{V}_{DaD} = g\Gamma_3\gamma^{\mu}\tau^{a}, \qquad \mathcal{V}_{DaD} = \begin{pmatrix} g\ \gamma^{\mu}\tau^{a} & 0 \\ 0 & -g\ \gamma^{\mu}\tau^{a} \end{pmatrix} = \begin{pmatrix} \mathcal{V}_{D} & 0 \\ 0 & \mathcal{V}_{aD} \end{pmatrix}. \tag{5.22}$$

Putting everything together we obtain the bi-spinor Feynman rules for the fermions of different scalar spin values. They are listed in Appendix D.

## 6. Scalar Spin of Bi-Spinors

In this section we will derive scalar spin of Dirac/anti-Dirac particles from the Lorentz transformation law of bi-spinors. Note that in this section the generation index $A$ runs from one to four. To obtain the operator of angular momentum consider transformation of bi-spinors under Lorentz transformation $\Lambda = \left\{\tilde{\Lambda}^{\mu}{}_{\nu}\right\} \in SO(1,3)$ with infinitesimal parameters $\delta\omega_{\mu\nu}$

$$\begin{gathered} \Psi(x) \to \Psi'(x') = S(\Lambda)\Psi\left(\tilde{\Lambda}^{-1}x\right)S^{-1}(\Lambda) = \Psi(x) + \delta\Psi(x), \quad \Lambda_{\mu\nu} = g_{\mu\sigma}\tilde{\Lambda}^{\sigma}{}_{\nu}, \\ \delta\Psi(x) = -\frac{i}{2}\delta\omega_{\mu\nu}\left(\left[\sigma^{\mu\nu}, \Psi(x)\right] + i\left(x^{\mu}\partial^{\nu} - x^{\nu}\partial^{\mu}\right)\Psi(x)\right) = -\frac{i}{2}\delta\omega_{\mu\nu}\hat{L}^{\mu\nu}\Psi(x), \end{gathered} \tag{6.1}$$

where the parameters of the Lorentz transformation are given by

$$\Lambda_{\rho\sigma} = g_{\rho\sigma} - \frac{i}{2}\delta\omega_{\mu\nu}M^{\mu\nu}{}_{\rho\sigma}, \quad M^{\mu\nu} = \left\{\left(M^{\mu\nu}\right)_{\rho\sigma}\right\}, \quad M^{\mu\nu}{}_{\rho\sigma} = i\left(\delta^{\mu}{}_{\rho}\delta^{\nu}{}_{\sigma} - \delta^{\mu}{}_{\sigma}\delta^{\nu}{}_{\rho}\right), \tag{6.2}$$

while the representation of the Lorentz group in the spinor space is given by

$$S(\Lambda)=\exp\left(-\frac{i}{2}\delta\omega_{\mu\nu}\sigma^{\mu\nu}\right)=1-\frac{i}{2}\delta\omega_{\mu\nu}\sigma^{\mu\nu}, \qquad \sigma^{\mu\nu}=\frac{i}{4}\left[\gamma^{\mu},\gamma^{\nu}\right]. \tag{6.3}$$

The exponential form of expression (6.3) for $S(\Lambda)$ is also valid when $\delta\omega_{\mu\nu}=1/2\left(\Lambda_{\mu\nu}-\Lambda_{\nu\mu}\right)$ is finite. It is easy to check that for finite values of $\delta\omega_{\mu\nu}$

$$\gamma^{0}S^{+}(\Lambda)\gamma^{0}S(\Lambda)=1. \tag{6.4}$$

In the Dirac $\gamma-$ matrix representation (6.4) means that $S(\Lambda)\in U(2,2)$. Therefore, (6.4) defines a mapping of the 6-parameter Lorentz group into 16-parameter group $U(2,2)$: $L\to U(2,2)$. Given an element of $U(2,2)$ in the image of the mapping, an element of $L$ can be reconstructed in a unique way

$$\Lambda_{\mu\nu}=g_{\mu\nu}+\mathrm{Re}\left(tr\left[\sigma_{\mu\nu}\ln S(\Lambda)\right]\right), \tag{6.5}$$

where we used $tr\left(\sigma^{\mu\nu}\sigma^{\rho\sigma}\right)=g^{\mu\rho}g^{\nu\sigma}-g^{\mu\sigma}g^{\nu\rho}$ and the logarithm is defined via Taylor expansion of matrix $S(\Lambda)$. Alternatively we can define the inverse mapping $U(2,2)\to L$ as the set of all transformations from $U(2,2)$, such that matrix $\Lambda_{\mu\nu}$

$$\Lambda_{\mu\nu}=g_{\mu\nu}+\mathrm{Re}\left(tr\left[\sigma_{\mu\nu}\ln U\right]\right), \tag{6.6}$$

satisfies

$$\Lambda_{\mu\nu}g^{\nu\rho}\Lambda_{\rho\sigma}=g_{\mu\sigma}. \tag{6.7}$$

Note that all other $\gamma-$ matrix representations are related to the Dirac representation by a similarity transformation and, therefore in other representations $S(\Lambda)$ belong to a subgroup of a group all complex matrices that preserves a bilinear form that is a similarity transformation of $\Gamma$ and is isomorphic to $U(2,2)$. Therefore, by choosing the Dirac representation we do not loose generality.

Like in the Dirac case bi-spinor angular momentum operator $\hat{L}^{\mu\nu}$

$$\hat{L}^{\mu\nu}=\left[\sigma^{\mu\nu},\cdot\ \right]+i\left(x^{\mu}\partial^{\nu}-x^{\nu}\partial^{\mu}\right), \tag{6.8}$$

contains intrinsic and orbital parts. The orbital part is identical to that of Dirac angular momentum operator $L^{\mu\nu}_{orb}=i\left(x^{\mu}\partial^{\nu}-x^{\nu}\partial^{\mu}\right)$. However, the intrinsic part in addition to the Dirac term $L^{\mu\nu}_{\mathrm{int}}=\sigma^{\mu\nu}\Psi(x)$ has an additional intrinsic term $\tilde{L}^{\mu\nu}_{\mathrm{int}}=-\Psi(x)\sigma^{\mu\nu}$, corresponding to the additional spinor index $\beta$ of bi-spinor $\Psi(x)=\left\{\Psi_{\alpha\beta}(x)\right\}$. Extracting from variation of massive bi-spinor action

$$S=\int d^{4}x\left(tr\left[\overline{\overline{\Psi}}(x)(i\partial\!\!\!/)\Psi(x)\right]-tr\left[\overline{\overline{M}}\ \overline{\overline{\Psi}}(x)\Psi(x)\right]-tr\left[\overline{\overline{\Psi}}(x)\Psi(x)M\right]\right), \tag{6.9}$$

the part that depends on derivatives of $\delta\omega_{\mu\nu}$ we obtain conserved bi-spinor angular momentum current density

$$J^{\mu,\rho\sigma} = \frac{1}{2} tr\left[\overline{\overline{\Psi}}\left(i\left(x^{\rho}\overleftrightarrow{\partial}^{\sigma} - x^{\rho}\overleftrightarrow{\partial}^{\sigma}\right)\gamma^{\mu}\right)\Psi\right] + \frac{1}{2} tr\left[\overline{\overline{\Psi}}\left\{\gamma^{\mu}, \sigma^{\rho\sigma}\right\}\Psi\right] - \frac{1}{2} tr\left[\overline{\overline{\Psi}}\,\gamma^{\mu}\,\Psi\sigma^{\rho\sigma}\right],$$

$$\partial_{\mu} J^{\mu,\rho\sigma} = 0 . \tag{6.10}$$

While the first two terms in (6.10) are identical to those for the Dirac case, the last term in (6.10) is specifically bi-spinor term. From (6.10) we obtain the conserved angular momentum tensor $J^{\mu\nu}$ as the space integral over the zero component of the density $J^{0,\rho\sigma}$

$$J^{\mu\nu} = \int d^{3}x\, J^{0,\rho\sigma}(x). \tag{6.11}$$

Current (6.10) is the basis of the conventional treatment of angular momentum in the bi-spinor gauge theory. Note, however, that (6.9) is invariant with respect to a symmetry group which larger than (6.1). It consists of transformations

$$\Psi(x) \to \Psi'(x) = S(\Lambda)\Psi\left(\Lambda^{-1}x\right)S^{-1}(\Lambda') = \Psi(x) + \delta\tilde{\Psi}(x),$$

$$M \to M' = S(\Lambda')M\, S^{-1}(\Lambda), \tag{6.12}$$

of which transformations in (6.1) form a diagonal subgroup. In (6.12) the left and the right transformations of $\Psi(x)$ are completely independent. If we use spinbein decomposition of $\Psi(x)$ with Dirac spinors $\xi^{A}(x)$ and constant spinbein $\eta^{A}$

$$\Psi_{\alpha\beta}(x) = \xi_{\alpha}{}^{A}(x)\,\overline{\overline{\eta}}^{A}{}_{\beta}, \tag{6.13}$$

(6.12) implies that under (6.12) $\xi^{A}(x)$ and $\eta^{A}$ transform independently. Bi-spinor $\Psi_{\alpha\beta}(x)$ is invariant under global $U(2,2)$ transformations (2.19) in the generation space. Under infinitesimal Lorentz transformation $\Lambda$ in (6.5) spinbein $\eta^{A}$ and Dirac spinors $\xi^{A}$ undergo transformations

$$\delta\xi^{A}(x) = -\frac{i}{2}\delta\omega_{\mu\nu}\left(\sigma^{\mu\nu}\xi^{A}(x) + i\left(x^{\mu}\partial^{\nu} - x^{\nu}\partial^{\mu}\right)\xi^{A}(x)\right) = -\frac{i}{2}\delta\omega_{\mu\nu}L^{\mu\nu}\xi^{A}(x),$$

$$\delta\eta^{A} = -\frac{i}{2}\delta\tilde{\omega}_{\mu\nu}\sigma^{\mu\nu}\eta^{A}. \tag{6.14}$$

Equation (6.7) implies that bi-spinors possess two types of angular momentum: the standard Dirac spinor angular momentum that has both orbital and intrinsic part and an additional purely intrinsic angular momentum.

However, this is not the end of story. Since spinbein $\eta^{A}$ indirectly defines the vacuum of quantum bi-spinor theory, it has to stay unchanged. Different vacuums defined by different spinbeins define inequivalent quantum field theories. They are inequivalent in the sense that, in general, there does not exist a unitary transformation that transforms one Fock space into another. Therefore, spinbein $\eta^A$ must remain unchanged. Alternatively, one can argue that by eliminating a spinbein in transition from bi-spinor formulation of the theory to its Dirac spinor formulation one has to make sure that all transformations affecting spinbein are carried over in their action on the corresponding multiplet of Dirac spinors. Therefore, spinbein $\eta^A$ must remain unchanged for that reason as well. One can implement constancy of spinbein under (6.5) by using $U(2,2)$ invariance (2.19) of the definition of $\Psi_{\alpha\beta}(x)$. Namely we will define a $U(2,2)$ matrix $\Omega = \left\{ \Omega^{AB}(\Lambda,\eta) \right\}$ by requiring that

$$\overline{\overline{\eta}} S^{-1}(\Lambda) = \Omega^T \overline{\overline{\eta}} \,. \tag{6.15}$$

Since $\overline{\overline{\eta}} = \Gamma \eta^+ \gamma^0$ we obtain

$$\Omega^T = \overline{\overline{\eta}} S^{-1}(\Lambda) \eta \,. \tag{6.16}$$

Obviously, because of spinbein normalization (2.13-14), the set of all $\Omega$ form a group. Since in the Dirac $\gamma-$ matrix representation $\eta, \overline{\overline{\eta}} \in U(2,2)$, mapping (6.16) defines a homomorphism of $L$ into $U(2,2) \cong SO(2,4)$.

We obtain in the end that the requirement that the spinbein remains invariant under Lorentz transformations of bi-spinors results a representation of Lorentz group $L$ in the generation space

$$\begin{aligned} &\delta \xi^A(x) = \left( -\frac{i}{2} \delta\omega_{\mu\nu} L^{\mu\nu} \delta^{AB} - \frac{i}{2} \delta\tilde{\omega}_{\mu\nu} \Sigma^{AB,\mu\nu} \right) \xi^B(x) \,, \\ &\delta \eta^A = 0 \,, \end{aligned} \tag{6.17}$$

where $L^{\mu\nu}$ is defined in (6.7) and

$$\Sigma^{AB,\mu\nu} = -\overline{\overline{\eta}}^B \sigma^{\mu\nu} \eta^A \,. \tag{6.18}$$

Thus bi-spinor theory has two independent conserved angular momentum current densities. The standard Dirac spinor angular momentum current density $J_D{}^{\mu,\rho\sigma}$ and an additional current density $J_S{}^{\mu,\rho\sigma}$

$$\begin{aligned} &J_D{}^{\mu,\rho\sigma} = \frac{1}{2} tr\left[ \overline{\overline{\psi}}^A \left( i \left( x^\rho \vec{\partial}^\sigma - x^\rho \vec{\partial}^\sigma \right) \gamma^\mu \right) \psi^A \right] + \frac{1}{2} tr\left[ \overline{\overline{\psi}}^A \left\{ \gamma^\mu, \sigma^{\rho\sigma} \right\} \psi^A \right], \\ &\partial_\mu J_D{}^{\mu,\rho\sigma} = 0 \end{aligned} \tag{6.19}$$

$$\begin{aligned} &J_S{}^{\mu,\rho\sigma} = -tr\left[ \overline{\overline{\psi}}^A \gamma^\mu \Sigma^{AB,\rho\sigma} \psi^B \right], \qquad \Sigma^{AB,\rho\sigma} = \overline{\overline{\eta}}^B \sigma^{\rho\sigma} \eta^A \,, \\ &\partial_\mu J_S{}^{\mu,\rho\sigma} = 0 \,. \end{aligned} \tag{6.20}$$

Current density $J_S{}^{\mu,\rho\sigma}$ is the origin of what we will call scalar spin quantum number, referred to in [10] as the right spin. Its existence is due solely to the invariance of spinbein decomposition (6.6) under the $U(2,2)$ transformations and to the existence of symmetry (6.5). We will show in Appendix A in more detail why we can call the corresponding quantum number the scalar spin quantum number. To finish the discussion of scalar spin we list two Pauli-Lubanski co-vectors in bi-spinor theory that describe intrinsic angular momentum. The classical Pauli-Lubanski co-vector $W_\mu$ is given by

$$W_\mu = -\frac{1}{2}\varepsilon_{\mu\nu\rho\sigma} J^{\nu\rho} P^\sigma , \tag{6.21}$$

where angular momentum tensor $J^{\nu\rho}$ is given by (6.10). Its bi-spinor analog $W_{S\mu}$ is given by the same expression

$$W_{S\mu} = -\frac{1}{2}\varepsilon_{\mu\nu\rho\sigma} J_S{}^{\nu\rho} P^\sigma , \tag{6.22}$$

but where the scalar spin angular momentum tensor $J_S{}^{\nu\rho}$ is defined by (6.20) and

$$J_S{}^{\mu\nu} = \int d^3x \, J_S{}^{0,\rho\sigma}(x). \tag{6.23}$$

## 7. Summary

In summary, we developed a formal perturbation theory for fermionic bi-spinors in bi-spinor gauge theory. The main difference with thee standard gauge theory is that a bi-spinor gauge theory contains global non-compact $U(2,2)$ symmetry. This introduces non-trivial changes into classification of elementary excitations and in the form of their propagators. In addition to the standard Dirac spinors elementary particles in bi-spinor gauge theory there also exist anti-Dirac spinors or Dirac-anti-Dirac generation space doublets. The appearance of the additional excitations can be traced to the additional non-dynamical spin, called scalar spin, that bi-spinors exhibit in their spinbein decompositions in the physical gauge.

In addition, we derived all possible mass terms for massive fermions in bi-spinor gauge theory. The solutions are classified by the scalar spin quantum number, a number that has no analog in the standard gauge theory. The possible mass terms correspond to combinations of scalar spin zero and 1/2 singlets and doublets in the generation space.

A description of the connection between Lorentz spin of bi-spinors and Lorentz and scalar spin of bi-spinor Dirac/anti-Dirac constituents was given, that shows how scalar spin for the algebraic Dirac constituents of bi-spinors arises from bi-spinors in the physical gauge.

It is the Dirac spinors rather then bi-spinors that are the mathematical objects used in the Standard Model to describe fermions. In this together with our previous work [21, 22, 23] we showed that the use of bi-spinors instead of Dirac spinors can bring certain advantages when describing quantum interactions of fermionic matter. The bi-spinor gauge theory is a renormalizable theory that allows one to avoid the use of torsion when describing coupling of fermions to gravity and leads to unique textures of lepton and quark mixing without introduction of additional degrees of freedom. All these features of bi-spinor gauge theory indicate that bi-spinors may offer a more fitting description of quantum fermionic matter. Of course, in the final count the description could only be better if the bi-spinor analog of the Standard Model generates better fit to the experimental data then the SM. We will consider this issue in a future publication.

## Appendix A: DaD Plane Wave Solutions

In this appendix we will determine plane wave solutions for equations of motion (5.6) for the scalar spin 1/2 DaD doublet. We look for solutions in the form $\xi(x) = \xi^{(\pm)}(k)e^{\mp ikx}$ for positive/negative energy. After substitution of $\xi(x)$ into (5.6) we obtain

$$\left(\pm \not{k} - m\left(c_\lambda + (-1)^A s_\lambda\, \gamma^5\right)\right)\xi^{(\pm)A}(k) = 0\,, \qquad A = 1,2\,, \tag{A.1}$$

where $c_\lambda = \cosh\lambda,\ s_\lambda = \sinh\lambda$ are the mass scaling parameters. We only need to solve for $\xi^1 \equiv \xi(k, s_\lambda)$ . $A = 2$ solutions are obtained by inverting the sign of $s_\lambda$ : $\xi^2 = \xi(k, -s_\lambda)$ . Now note that on-shell with $\left(k^2 - m^2\right)\xi^A = 0$ we have

$$\left(\pm \not{k} - m\left(c_\lambda - s_\lambda\, \gamma^5\right)\right)\left(\pm \not{k} + m\left(c_\lambda + s_\lambda\, \gamma^5\right)\right) = 0\,, \tag{A.2}$$

Therefore, solutions of (A.1) for $A = 1$ are given by

$$\xi^{(\pm)}\left(\vec{k}, s_\lambda\right) = \left(\pm \not{k} + m\left(c_\lambda + s_\lambda\, \gamma^5\right)\right)\hat{\xi}^{(\pm)}, \tag{A.3}$$

where $\hat{\xi}^{(\pm)}$ are four-component spinors that depend on two parameters only, because $\left(\pm \not{k} + m\left(c_\lambda + s_\lambda\, \gamma^5\right)\right)$ is matrix of rank two. The independent components of $\hat{\xi}^{(\pm)}$ can be determined from the solutions in the $k^\mu = \left(m, \vec{0}\right)$ rest frame. Using Dirac $\gamma$ -matrix representation with

$$\gamma^0 = \begin{pmatrix} \mathbf{1} & 0 \\ 0 & -\mathbf{1} \end{pmatrix}, \quad \gamma^k = \begin{pmatrix} 0 & \sigma^k \\ -\sigma^k & 0 \end{pmatrix}, \quad \gamma^5 = \begin{pmatrix} 0 & \mathbf{1} \\ \mathbf{1} & 0 \end{pmatrix},$$

and metric convention $g_{\mu\nu} = diag(1, -1, -1, -1)$ we obtain for the rest frame solutions: $\left(\pm\gamma^0 - c_\lambda + s_\lambda\, \gamma^5\right)\xi^{(\pm)}(m, s_\lambda) = 0$ , or

$$\begin{pmatrix} \pm 1 - c_\lambda & s_\lambda \\ s_\lambda & \mp 1 - c_\lambda \end{pmatrix}\begin{pmatrix} \varsigma^1 \\ \tilde{\varsigma}^1 \end{pmatrix} = 0\,. \tag{A.4}$$

From (A.4) we obtain positive and negative energy solutions for $\xi^{(\pm)}(m, s_\lambda)$

$$\begin{aligned} \xi_r^{(+)}(m, s_\lambda) &= \begin{pmatrix} \varsigma_r \\ \dfrac{s_\lambda}{(1 + c_\lambda)}\varsigma_r \end{pmatrix} = (1 + c_\lambda)^{-1}\left(+\gamma^0 + c_\lambda + s_\lambda\, \gamma^5\right)\begin{pmatrix} \varsigma_r \\ 0 \end{pmatrix}, \\ \xi_r^{(-)}(m, s_\lambda) &= \begin{pmatrix} \dfrac{s_\lambda}{(1 + c_\lambda)}\eta_r \\ \eta_r \end{pmatrix} = (1 + c_\lambda)^{-1}\left(-\gamma^0 + c_\lambda + s_\lambda\, \gamma^5\right)\begin{pmatrix} 0 \\ \eta_r \end{pmatrix}, \end{aligned} \tag{A.5}$$

where $\varsigma_r$, $\eta_r$, $r=1,2$, form two bases in the two-component spinor space. Combining (A.3) with (A.5-6) we obtain the $+s$ basis positive/negative energy solutions in (5.6)

$$\xi_r^{(+)}\left(\vec{k},s_\lambda\right)=N\left(\vec{k}\right)\begin{pmatrix}\varsigma_r\\ \left(\dfrac{\vec{\sigma}\,\vec{k}+ms_\lambda}{E_k+mc_\lambda}\right)\varsigma_r\end{pmatrix},$$

$$\xi_r^{(-)}\left(\vec{k},s_\lambda\right)=N\left(\vec{k}\right)\begin{pmatrix}\left(\dfrac{\vec{\sigma}\,\vec{k}+ms_\lambda}{E_k+mc_\lambda}\right)\eta_r\\ \eta_r\end{pmatrix},\qquad E_k\equiv k^0=+\sqrt{\vec{k}^2+m^2}\,, \tag{A.6}$$

where $N\left(\vec{k}\right)$ is a normalization factor determined by normalization of the energy-momentum vector $P^\mu$.

$$N\left(\vec{k}\right)=\left[\frac{m}{E_k}\left(1+\frac{\vec{k}^2-m^2{s_\lambda}^2}{\left(E_k+mc_\lambda\right)^2}\right)\right]^{-\frac{1}{2}}. \tag{A.7}$$

Dirac contractions of the solutions are given by

$$\bar{\xi}_r^{(+)}\left(\vec{k},s_\lambda\right)\xi_{r'}^{(+)}\left(\vec{k},s_\lambda\right)=+\left|N\left(\vec{k}\right)\right|^2\left[\left(1-\frac{\vec{k}^2+m^2{s_\lambda}^2}{\left(E_k+mc_\lambda\right)^2}\right)\delta_{rr'}-\frac{2ms_\lambda}{\left(E_k+mc_\lambda\right)^2}\left(\vec{\sigma}\cdot\vec{k}\right)_{rr'}\right],$$

$$\bar{\xi}_r^{(-)}\left(\vec{k},s_\lambda\right)\xi_{r'}^{(-)}\left(\vec{k},s_\lambda\right)=-\left|N\left(\vec{k}\right)\right|^2\left[\left(1-\frac{\vec{k}^2+m^2{s_\lambda}^2}{\left(E_k+mc_\lambda\right)^2}\right)\delta_{rr'}-\frac{2ms_\lambda}{\left(E_k+mc_\lambda\right)^2}\left(\vec{\sigma}\cdot\vec{k}\right)_{rr'}\right], \tag{A.8}$$

$$\bar{\xi}_r^{(+)}\left(\vec{k},s_\lambda\right)\xi_{r'}^{(-)}\left(\vec{k},s_\lambda\right)=0\,.$$

We can now determine the positive/negative modes entering in (5.13)

$$\begin{pmatrix}u_{r1}^1\left(\vec{k}\right)\\ u_{r1}^2\left(\vec{k}\right)\end{pmatrix}=N\left(\vec{k}\right)\begin{pmatrix}0\\ \dfrac{ms_\lambda}{E_k+mc_\lambda}\varsigma_r\\ -\varsigma_r\\ -\dfrac{\vec{\sigma}\,\vec{k}}{E_k+mc_\lambda}\varsigma_r\end{pmatrix},\qquad \begin{pmatrix}u_{r2}^1\left(\vec{k}\right)\\ u_{r2}^2\left(\vec{k}\right)\end{pmatrix}=N\left(\vec{k}\right)\begin{pmatrix}\varsigma_r\\ \dfrac{\vec{\sigma}\,\vec{k}}{E_k+mc_\lambda}\varsigma_r\\ 0\\ -\dfrac{ms_\lambda}{E_k+mc_\lambda}\varsigma_r\end{pmatrix}, \tag{A.9}$$

$$\begin{pmatrix} v^1_{r1}(\vec{k}) \\ v^2_{r1}(\vec{k}) \end{pmatrix} = N(\vec{k}) \begin{pmatrix} \frac{m s_\lambda}{E_k + m c_\lambda} \eta_r \\ 0 \\ -\frac{\vec{\sigma}\,\vec{k}}{E_k + m c_\lambda} \eta_r \\ -\eta_r \end{pmatrix}, \qquad \begin{pmatrix} v^1_{r2}(\vec{k}) \\ v^2_{r2}(\vec{k}) \end{pmatrix} = N(\vec{k}) \begin{pmatrix} \frac{\vec{\sigma}\,\vec{k}}{E_k + m c_\lambda} \eta_r \\ \eta_r \\ \frac{m s_\lambda}{E_k + m c_\lambda} \eta_r \\ 0 \end{pmatrix}.$$

Their contractions are given by

$$\begin{aligned} u^A{}_{rp}{}^+(\vec{k})\, u^A{}_{r'q}(\vec{k}) &= \frac{2E_k}{E_k + mc} \delta_{rr'} \delta_{pq}, \\ v^A{}_{rp}{}^+(\vec{k})\, v^A{}_{r'q}(\vec{k}) &= \frac{2E_k}{E_k + mc} \delta_{rr'} \delta_{pq}, \\ u^A{}_{rp}{}^+(\vec{k})\, v^A{}_{r'q}(-\vec{k}) &= 0. \end{aligned} \tag{A.10}$$

## Appendix B: Diagonalization of the Hamiltonian bilinears

In this Appendix we describe diagonalization of the DaD Hamiltonian to show how the $O(2)$ rotation $T_k$ appears in (5.9). From the Lagrangian density

$$\mathcal{L}_{DaD} = \overline{\overline{\psi}}(i\partial\!\!\!/ - m c_\lambda)\psi - m s_\lambda\, \overline{\overline{\psi}}\, \Gamma_2 \gamma^5 \psi, \qquad \overline{\overline{\psi}} = \overline{\psi}\, \Gamma_3. \tag{B.1}$$

we obtained the conjugate momenta, which are defined by

$$\pi^A = \frac{\partial \mathcal{L}_{DaD}}{\overleftarrow{\partial}(\partial_0 \psi^A)} = i \overline{\overline{\psi}}{}^A \gamma^0, \qquad A = 1,2, \tag{B.2}$$

where the arrow indicates from which side the anti-commuting derivative acts. The Hamiltonian density is obtained using the Legendre transformation

$$\mathcal{H}_{DaD} = \pi^A \partial_0 \psi^A - \mathcal{L}_{DaD} = tr\left[\overline{\overline{\psi}}\left(i\vec{\gamma}\cdot\vec{\nabla} + m c_\lambda\right)\psi + m s_\lambda\, \overline{\overline{\psi}}\, \gamma^5 \psi\right], \tag{B.3}$$

where trace is over the Dirac and the generation indices. From (B.1) we obtain equations of motion

$$\begin{aligned} (i\partial\!\!\!/ - m c_\lambda)\psi^1 - m s_\lambda \gamma^5 \psi^2 &= 0, \\ (i\partial\!\!\!/ - m c_\lambda)\psi^2 - m s_\lambda \gamma^5 \psi^1 &= 0. \end{aligned} \tag{B.4}$$

After substitution of the equations of motion the Hamiltonian density for DaD doublet in the new field variables (5.4) becomes

$$\mathcal{H} = \bar{\xi}^2 (i\gamma^0 \partial_0) \xi^1 + c.c.. \tag{B.5}$$

We now can write out this Hamiltonian in terms of the positive/negative energy modes. First, to show how indefiniteness of DaD Hamiltonian manifests itself under standard quantization rules, we will use the standard creation-annihilation operator assignment. We obtain the expansion for a 8-component of DaD doublet field

$$\xi(x) = \int \frac{d^3k}{(2\pi)^3} \frac{m}{k^0} \sum_{r,p} \left[ \hat{\xi}^{(+)}_{rp}(\vec{k})\, \hat{b}_{rp}(\vec{k})\, e^{-ikx} + \hat{\xi}^{(-)}_{rp}(\vec{k})\, \hat{d}_{rp}{}^{+}(\vec{k})\, e^{ikx} \right]. \quad \text{(B.6)}$$

We now substitute (A.6) into the expression for the Hamiltonian $H = \int d^3x\, \mathcal{H}$ and obtain

$$H = \int d^3k \left( \hat{b}_{r2}{}^{+}\hat{b}_{r1} + \hat{b}_{r1}{}^{+}\hat{b}_{r2} + \hat{d}_{r2}{}^{+}\hat{d}_{r1} + \hat{d}_{r1}{}^{+}\hat{d}_{r2} \right). \quad \text{(B.7)}$$

This expression is diagonalized by a $O(2)$ rotation that is in the opposite direction of rotation $T_x$ in (5.9)

$$\begin{pmatrix} \hat{b}_{r1} \\ \hat{b}_{r2} \end{pmatrix} = \frac{1}{\sqrt{2}} \begin{pmatrix} 1 & -1 \\ 1 & 1 \end{pmatrix} \begin{pmatrix} b_{r1} \\ b_{r2} \end{pmatrix}, \qquad \begin{pmatrix} \hat{d}_{r1} \\ \hat{d}_{r2} \end{pmatrix} = \frac{1}{\sqrt{2}} \begin{pmatrix} 1 & -1 \\ 1 & 1 \end{pmatrix} \begin{pmatrix} d_{r1} \\ d_{r2} \end{pmatrix}. \quad \text{(B.8)}$$

We recognize in (B.8) the $O(2)$ rotation $T_k$ in (5.9). Substitution of (B.8) into (B.7) after normal ordering results in

$$H = \int \frac{d^3k}{(2\pi)^3}\, m \left( b_{r1}{}^{+} b_{r1} - b_{r2}{}^{+} b_{r2} + d_{r1}{}^{+} d_{r1} - d_{r2}{}^{+} d_{r2} \right). \quad \text{(B.9)}$$

As expected the use of the standard Dirac assignment to creation-annihilation operators results in a Hamiltonian that is not bounded from below. To cure the problem we must use the flipped Dirac operator assignments by making a replacement

$$b_{r2}{}^{+} \leftrightarrow b_{r2}, \quad d_{r2}{}^{+} \leftrightarrow d_{r2}\,. \quad \text{(B.10)}$$

After reassignment (B.10) we obtain the final form of the normalized plane-wave solutions in (5.7).

## Appendix C: Scalar Spin States in Fock Space

In this Appendix we will establish a precise meaning of the scalar spin quantum number, which appears as an additional index in the DaD creation/annihilation operators in the bi-spinor gauge theory. To make the analogy between the Lorentz and scalar spin more transparent we will us a degenerate DaD scalar spin doublet, where the Dirac and the Anti-Dirac components have the same mass. To understand the physical meaning of indexes in $b_{rp}, d_{r'q}$ let us consider one-particle states in the bi-spinor Fock space. Following the standard exposition in we consider the states with $\left|b_{rp}\right\rangle = b_{rp}^{+}\left|0\right\rangle$ and $\left|d_{rp}\right\rangle = d_{rp}^{+}\left|0\right\rangle,\ r,p = 1,2$ . From (5.7, 5.15) we obtain

$$Q\left|b_{rp}\right\rangle = +\left|b_{rp}\right\rangle, \qquad Q\left|d_{rp}\right\rangle = -\left|d_{rp}\right\rangle. \tag{C.1}$$

Therefore, states $\left|b_{rp}\right\rangle$ have definite positive charge, while $\left|d_{rp}\right\rangle$ have negative charge. We now consider the action of angular momentum and scalar angular momentum operators on the states $\left|b_{rp}\right\rangle$ and $\left|d_{rp}\right\rangle$.

To proceed further we note that from (5.7) we obtain

$$\begin{aligned} b_{r1}(\vec{k}) &= \int d^3x \overline{u}^A_{r1}(\vec{k}) \exp(i k x) \gamma^0 \psi^A(x), & b_{r2}{}^{+}(\vec{k}) &= \int d^3x \overline{u}^A_{r2}(\vec{k}) \exp(i k x) \gamma^0 \psi^A(x), \\ d_{r1}(\vec{k}) &= \int d^3x \overline{v}^A_{r1}(\vec{k}) \exp(-i k x) \gamma^0 \psi^A(x), & d_{r2}(\vec{k}) &= \int d^3x \overline{v}^A_{r2}(\vec{k}) \exp(i k x) \gamma^0 \psi^A(x). \end{aligned} \tag{C.2}$$

Our next step us to evaluate the commutators of the operators of Lorentz and scalar spin with $b_{rp}{}^{+}, d_{rp}{}^{+}$.

First, we need to discuss the relation between the quadruplet $\psi^{\overline{A}}(x),\ \overline{A} = 1,\ldots,4$ that we have used in Section 5 and the DaD doublets $\psi^{A}(x),\ A = 1,2$ , we have considered in this Appendix. Recall that $\psi^{\overline{A}}(x)$ appear as the result of spinbein decomposition (2.15) for $U(1)$ gauge group or (2.21) for non-Abelian gauge group of bi-spinor $\Psi_{\alpha\beta}(x)$ with a constant spinbein $\eta^{\overline{A}}_{\beta}$

$$\Psi_{\alpha\beta}(x) = \xi^{\overline{A}}_{\alpha}(x) \overline{\eta}^{\overline{A}}_{\beta}, \qquad \overline{\eta} = \Gamma \eta^{+} \gamma^0, \tag{2.15}$$

where $\eta^{\overline{A}}_{\beta}$ is normalized according to (2.14) or (2.22). For simplicity we will consider only the $U(1)$ with

$$\eta^{\overline{A}}_{\alpha} \overline{\eta}^{\overline{A}}_{\beta} = \delta_{\alpha\beta}. \tag{2.14}$$

Note now that in the Dirac representation $\gamma^0 = diag(1,\ 1, -1 -1) = \Gamma$ . Consequently, we obtain (2.17), which implies that then spinbeins belong to $U(2,2)$. Therefore, we can represent an arbitrary spinbein matrix $\eta_{\alpha}{}^{\overline{A}}$ via its Cartan decomposition as a product of two $U(4)$ matrices and a symmetric matrix

$$\eta = U\,S\,V\,, \qquad \eta_\alpha{}^{\bar A} = U_{\alpha\beta}\,S_\beta{}^{\bar B}\,V^{\bar B\bar A}\,, \qquad U, V \in U(4), \tag{C.3}$$

where the two-parameter matrix $S_\beta{}^{\bar B}$ is given by

$$S_\beta{}^{\bar B} = \begin{pmatrix} \cosh\lambda_1 & 0 & \sinh\lambda_1 & 0 \\ 0 & \cosh\lambda_2 & 0 & \sinh\lambda_2 \\ \sinh\lambda_1 & 0 & \cosh\lambda_1 & 0 \\ 0 & \sinh\lambda_2 & 0 & \cosh\lambda_2 \end{pmatrix} = \begin{pmatrix} \cosh\lambda_1 & 0 & \sinh\lambda_1 & 0 \\ 0 & 0 & 0 & 0 \\ \sinh\lambda_1 & 0 & \cosh\lambda_1 & 0 \\ 0 & 0 & 0 & 0 \end{pmatrix} + \begin{pmatrix} 0 & 0 & 0 & 0 \\ 0 & \cosh\lambda_2 & 0 & \sinh\lambda_2 \\ 0 & 0 & 0 & 0 \\ 0 & \sinh\lambda_2 & 0 & \cosh\lambda_2 \end{pmatrix}. \tag{C.4}$$

We recognize in the two summands the entries from admissible mass matrices for two DaD doublets $\mathcal{M}_R{}^{(2)}$ in (3.17). The first summand in (C.4) can be considered as a spinbein $\eta_\alpha^{\bar A}(\lambda_1)$ in its own right. The same applies to the second summand, denoted as $\eta_\alpha^{\bar A}(\lambda_2)$. The two are degenerate spinbeins with normalizations

$$\begin{aligned} \eta_\alpha^{\bar A}(\lambda_1)\overline{\overline{\eta}}_\beta{}^{\bar A}(\lambda_1) &= diag(1, 0, -1, 0), \\ \eta_\alpha^{\bar A}(\lambda_2)\overline{\overline{\eta}}_\beta{}^{\bar A}(\lambda_2) &= diag(0, 1, 0, -1). \end{aligned} \tag{C.5}$$

Using (C.5) an arbitrary spinbein $\eta_\alpha{}^{\bar A}$ can be expressed in terms of two degenerate spinbeins $\eta_\alpha^{\bar A}(\lambda_k)$ according to

$$\begin{aligned} \eta_\alpha{}^{\bar A} &= U_{\alpha\beta}\left(\eta_\beta^{\bar B}(\lambda_1) + \eta_\beta^{\bar B}(\lambda_2)\right)V^{\bar B\bar A}, \\ \eta &= U(\eta_1 + \eta_2)V\,, \qquad \eta_k = \eta(\lambda_k). \end{aligned} \tag{C.6}$$

We will call $\eta_k = \eta(\lambda_k)$ canonical degenerate spinbeins or simply canonical spinbeins. Now spinbein decomposition (2.15) can be written as

$$\Psi(x) = \xi(x)\overline{\overline{V}}\left(\overline{\overline{\eta}}_1 + \overline{\overline{\eta}}_2\right)\overline{\overline{U}}\,, \qquad \overline{\overline{U}} = \gamma^0 U^+ \gamma^0\,, \qquad \overline{\overline{V}} = \Gamma V^+ \Gamma\,. \tag{C.7}$$

We now choose such $U, V \in U(4)$ as to enable us to disentangle the two canonical spinbeins and the two DaD doublets implicitly contained in a bi-spinor. This can be done if $U$ commutes with $\gamma^0$ and $V$ commutes with $\Gamma$, which means that $U, V$
are block diagonal with each block a unitary $2\times 2$ matrix. We can now write

$$\Psi(x) = \Psi_1(x) + \Psi_2(x), \qquad \Psi_k(x) = \xi(x)V^+\overline{\overline{\eta}}_k\,U^+\,, \qquad k = 1,2\,. \tag{C.8}$$

Action of angular momentum operators on the quadruplet $\psi^{\bar A}(x)$, $\bar A = 1,\ldots,4$, is given by (6.14, 6.17, 6.19, 6.20)

$$\left[J_D{}^{\mu\nu}, \psi^{\bar A}(x)\right]_\alpha = -iL^{\mu\nu}{}_{\alpha\beta}\,\psi^{\bar A}{}_\beta(x),$$

$$L^{\mu\nu}{}_{\alpha\beta} = i\left(x^{\mu}\partial^{\nu} - x^{\nu}\partial^{\mu}\right)\delta_{\alpha\beta} + \left(\sigma^{\mu\nu}\right)_{\alpha\beta},$$

$$\left[J_S{}^{\mu\nu}, \psi^{\bar{A}}(x)\right] = -i\Sigma^{\bar{A}\bar{B},\mu\nu}\psi^{\bar{B}}(x), \tag{C.10}$$

$$\Sigma^{\bar{A}\bar{B},\mu\nu} = -\bar{\bar{\eta}}^{\bar{B}}\,\sigma^{\mu\nu}\eta^{\bar{A}}, \qquad \bar{A}, \bar{B} = 1,\ldots,4.$$

Since for the chosen spinbein $\Sigma^{\bar{A}\bar{B},\mu\nu}$ becomes block-diagonal, we can split the space of $\psi^{\bar{A}}(x)$ into two two-dimensional invariant subspaces spanned by $\psi_1{}^A(x)$, $\psi_2{}^A(x)$, $A = 1,2$. Without loss of generality we can now concentrate on one of the two subspaces. We obtain

$$\Sigma^{\bar{A}\bar{B},\mu\nu} = \begin{pmatrix} \Sigma_1{}^{AB,\mu\nu} & 0 \\ 0 & \Sigma_2{}^{AB,\mu\nu} \end{pmatrix}, \qquad \Sigma_k{}^{AB,\mu\nu} = \bar{\bar{\eta}}_k{}^B\sigma^{12}\eta_k{}^A,$$

$$(\eta_k)_\alpha{}^1 = \begin{pmatrix} c \\ s \end{pmatrix}, \qquad (\eta_k)_\alpha{}^2 = \begin{pmatrix} s \\ c \end{pmatrix}, \qquad (\eta_k)_\alpha{}^A = \begin{pmatrix} c & s \\ s & c \end{pmatrix}. \tag{C.9}$$

Dropping indexes referring to the subspaces we obtain for commutators of angular momentum tensor $J_D{}^{\mu\nu}$

$$\left[J_D{}^{\mu\nu}, b_{r1}{}^+\right] = \int d^3x\, \bar{\bar{\psi}}(x)\,\sigma^{\mu\nu}\,\gamma^0 u_{r1}(\vec{k})\exp(-ikx) + orbital\ contibution,$$

$$\left[J_D{}^{\mu\nu}, b_{r2}{}^+\right] = -\int d^3x\ \bar{\bar{u}}_{r2}(\vec{k})\gamma^0\ \sigma^{\mu\nu}\ \psi(x)\exp(+ikx) + orbital\ contibution,$$

$$\left[J_D{}^{\mu\nu}, d_{r1}{}^+\right] = -\int d^3x\ \bar{\bar{v}}_{r1}(\vec{k})\gamma^0\ \sigma^{\mu\nu}\ \psi(x)\exp(-ikx) + orbital\ contibution, \tag{C.11}$$

$$\left[J_D{}^{\mu\nu}, d_{r2}{}^+\right] = \int d^3x\, \bar{\bar{\psi}}(x)\,\sigma^{\mu\nu}\,\gamma^0 v_{r2}(\vec{k})\exp(+ikx) + orbital\ contibution,$$

where we made explicit only the intrinsic part of the commutators, since the orbital part cancels out when acting on the one-particle states. For commutators of scalar spin tensor $J_S{}^{\mu\nu}$ we obtain

$$\left[J_S{}^{\mu\nu}, b_{r1}{}^+\right] = -\int d^3x\, \bar{\bar{\psi}}(x)\frac{\Sigma^{\mu\nu}}{2}\gamma^0 u_{r1}(\vec{k})\exp(-ikx), \quad \Sigma^{\mu\nu} = \Sigma_1{}^{\mu\nu} = \left\{\Sigma^{AB,\mu\nu}\right\}$$

$$\left[J_S{}^{\mu\nu}, b_{r2}{}^+\right] = \int d^3x\ \bar{\bar{u}}_{r2}(\vec{k})\gamma^0\ \frac{\Sigma^{\mu\nu}}{2}\ \psi(x)\exp(+ikx),$$

$$\left[J_S{}^{\mu\nu}, d_{r1}{}^+\right] = \int d^3x\ \bar{\bar{v}}_{r1}(\vec{k})\gamma^0\ \frac{\Sigma^{\mu\nu}}{2}\ \psi(x)\exp(-ikx), \tag{C.12}$$

$$\left[J_S{}^{\mu\nu}, d_{r2}{}^+\right] = -\int d^3x\, \bar{\bar{\psi}}(x)\frac{\Sigma^{\mu\nu}}{2}\gamma^0 v_{r2}(\vec{k})\exp(+ikx).$$

Action of the Pauli-Lubanski operators (6.21-22) $(W\cdot n)/m$, $(W_S\cdot n)/m$, contracted with the unit vector $n^\mu$, which is chosen to be orthogonal to $k^\mu$,

$$n^\mu = \frac{m}{|\vec{k}|}\left(n_0^\mu - k^\mu \frac{k\,n_0}{m^2}\right), \qquad n_0^\mu = (1,0,0,0) \tag{C.13}$$

on the states $|b_{rp}\rangle$, $|d_{rp}\rangle$ with definite four-momentum $P^\sigma$ can be written as

$$\begin{aligned}
&\frac{1}{m}(W\cdot n)|b_{rp}\rangle = J_D{}^{12}|b_{rp}\rangle, \qquad && \frac{1}{m}(W\cdot n)|d_{rp}\rangle = J_D{}^{12}|d_{rp}\rangle,\\
&\frac{1}{m}(W_S\cdot n)|b_{rp}\rangle = J_S{}^{12}|b_{rp}\rangle, && \frac{1}{m}(W_S\cdot n)|d_{rp}\rangle = J_S{}^{12}|d_{rp}\rangle.
\end{aligned} \tag{C.14}$$

Since vacuum carries no angular momentum, we obtain

$$\begin{aligned}
\frac{1}{m}(W\cdot n)|b_{r1}\rangle &= \frac{1}{m}(W\cdot n)b_{r1}{}^{+}|0\rangle = +\frac{m}{k^0}u^{A+}{}_{sp}(\vec{k})\,\sigma^{12}u^{A}{}_{r1}(\vec{k})\,b_{sp}{}^{+}|0\rangle,\\
\frac{1}{m}(W\cdot n)|b_{r2}\rangle &= \frac{1}{m}(W\cdot n)b_{r2}{}^{+}|0\rangle = -\frac{m}{k^0}v^{A+}{}_{r2}(\vec{k})\,\sigma^{12}v^{A}{}_{sp}(\vec{k})\,b_{sp}{}^{+}|0\rangle,\\
\frac{1}{m}(W\cdot n)|d_{r1}\rangle &= \frac{1}{m}(W\cdot n)d_{r1}{}^{+}|0\rangle = -\frac{m}{k^0}v^{A+}{}_{r1}(\vec{k})\,\sigma^{12}v^{A}{}_{sp}(\vec{k})\,d_{sp}{}^{+}|0\rangle,\\
\frac{1}{m}(W\cdot n)|d_{r2}\rangle &= \frac{1}{m}(W\cdot n)d_{r2}{}^{+}|0\rangle = +\frac{m}{k^0}u^{A+}{}_{sp}(\vec{k})\,\sigma^{12}u^{A}{}_{r2}(\vec{k})\,d_{sp}{}^{+}|0\rangle,
\end{aligned} \tag{C.15}$$

where

$$\sigma^{12} = \begin{pmatrix} \frac{\sigma^3}{2} & 0 \\ 0 & \frac{\sigma^3}{2} \end{pmatrix}. \tag{C.16}$$

From (C.16) we can read off the Lorentz spin quantum number and the first of the two lower indices in $b_{rp}{}^{+}$, $d_{rp}{}^{+}$. For scalar spin we obtain

$$\begin{aligned}
\frac{1}{m}(W_S\cdot n)|b_{r1}\rangle &= \frac{1}{m}(W_S\cdot n)b_{r1}{}^{+}|0\rangle = -\frac{m}{k^0}u^{A+}{}_{sp}(\vec{k})\Sigma^{AB,12}u^{B}{}_{r1}(\vec{k})\,b_{sp}{}^{+}|0\rangle,\\
\frac{1}{m}(W_S\cdot n)|b_{r2}\rangle &= \frac{1}{m}(W_S\cdot n)b_{r2}{}^{+}|0\rangle = +\frac{m}{k^0}v^{A+}{}_{r2}(\vec{k})\Sigma^{AB,12}v^{B}{}_{sp}(\vec{k})\,b_{sp}{}^{+}|0\rangle,
\end{aligned} \tag{C.17}$$

$$\frac{1}{m}(W_S\cdot n)|d_{r1}\rangle=\frac{1}{m}(W_S\cdot n)d_{r1}{}^{+}|0\rangle=+\frac{m}{k^0}v^{A+}{}_{r1}(\vec{k})\Sigma^{AB,12}v^{B}{}_{sp}(\vec{k})\,d_{sp}{}^{+}|0\rangle,$$

$$\frac{1}{m}(W_S\cdot n)|d_{r2}\rangle=\frac{1}{m}(W_S\cdot n)d_{r2}{}^{+}|0\rangle=-\frac{m}{k^0}u^{A+}{}_{sp}(\vec{k})\Sigma^{AB,12}u^{B}{}_{r2}(\vec{k})\,d_{sp}{}^{+}|0\rangle,$$

where

$$\Sigma^{AB,12}=\overline{\overline{\eta}}^{B}\sigma^{12}\eta^{A},\ \eta=\{\eta_\alpha^A\}\in U(1,1). \qquad \text{(C.18)}$$

Substitution of the values for spinbein

$$\tilde{\eta}_\alpha{}^1=\begin{pmatrix}c\\ s\\ 0\\ 0\end{pmatrix}=\begin{pmatrix}\eta^1\\ 0\end{pmatrix},\qquad \tilde{\eta}_\alpha{}^2=\begin{pmatrix}s\\ c\\ 0\\ 0\end{pmatrix}=\begin{pmatrix}\eta^2\\ 0\end{pmatrix},\qquad \overline{\overline{\eta}}^{A}\eta^{B}=(-1)^{A-1}\delta^{AB}, \qquad \text{(C.19)}$$

into $\Sigma^{AB,12}$ results, after taking into account that $\overline{\overline{\tilde{\eta}}}^{A}=\begin{pmatrix}(\Gamma_3\eta)^A\\ 0\end{pmatrix}$, in

$$\{\Sigma^{AB,12}\}=\begin{pmatrix}\frac{1}{2} & 0\\ 0 & -\frac{1}{2}\end{pmatrix}. \qquad \text{(C.20)}$$

Note that the $4\times 4$ matrices (C.16, C.20) are acting in the space of irreducible two-component Dirac doublets. Comparing (C.20) with (C.16) we see that first four entries in the DaD octuplet $\psi^{\bar{A}}(x)$ describe a state with scalar spin $+1/2$, while the remaining four entries describe a state with scalar spin $-1/2$. This explains the meaning of the second lower index in $b_{rp}{}^{+}$, $d_{rp}{}^{+}$.

Let us now turn to the general case with non-degenerate DaD doublets. In such a case one cannot use the Pauli-Lubanski Casimir operator, because the elements of the DaD doublet have different mass. Instead one uses another, the more fitting Casimir operator, which is the square of $J_D{}^{AB}$, the scalar spin “angular momentum” in the $4\times 4$ generation space. One arrives to $J_D{}^{AB}$ as follows. A general bi-spinor transforms in the bi-fundamental representation, which simply means the both the Dirac spinor $\psi^{\bar{A}}(x)$ and the spinbein constituents $\eta^{\bar{A}}$ of the bi-spinor decomposition transform as Dirac spinors. We note that when spinbein undergoes a Lorentz transformation the Dirac spinors must not. Therefore, the Dirac spinor components in the bi-spinor decomposition relate to the old spinbein. A spinbein is a physical quantity, because it determines masses of the particles. Hence, the old Dirac spinors actually describe something different then before. To avoid this we need to make sure that the spinbein does not change under its Lorentz transformations. For constant spinbeins that we use the Lorentz transformation reduces to Lorentz rotations and they can be made unchanged with a compensatory rotation in the flavor space. This is possible because constant spinbeins are elements of $U(2,2)$. Recall now that $SO(1,3)\in SO(2,4)\cong U(2,2)$, where the second sign is the sign of isomorphism. Since there is one non-compact dimension in $SO(1,3)$ and two in $U(2,2)$ there are exactly two inequivalent ways to imbed $SO(1,3)$ into $U(2,2)$. Therefore, we can compensate the Lorentz rotation of the spinbein by using two different $\lambda_k$, $k = 1, 2$ in (C.4) in two different rotations in flavor space that are equivalent to flavor "Lorentz rotations". The representation theory of such an imbedded $SO(1, 3)$ in the flavor space is identical to the one in the Minkowski space. Therefore, we have exactly the same set of irreducible representations and their classification through a maximal set of the Casimir operators. Thus we obtain that the general $\psi^{\bar{A}}(x)$, which contains two DaD doublets, is described by two sets of scalar spins, one for each DaD doublet, each component of which has the value of the scalar spin $\pm 1/2$ as we observe in (C.20).

## Appendix D: Feynman Rules for Dirac and anti-Dirac Particles

Incoming and outgoing lines

Dirac spinors

Fig.2. Particle incoming line.

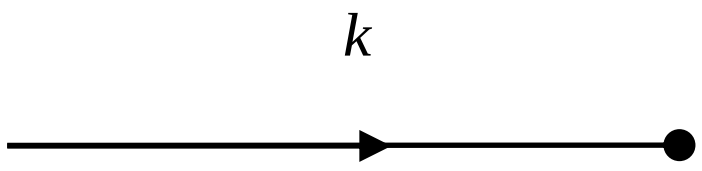

$u_\sigma(k)$

Fig. 3. Particle outgoing line.

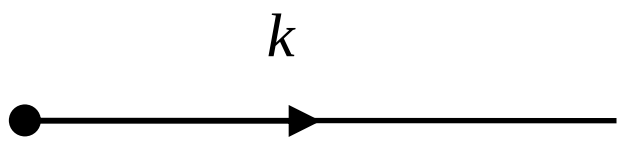

$\bar{u}_\sigma(k)$

Fig. 4. Antiparticle incoming line.

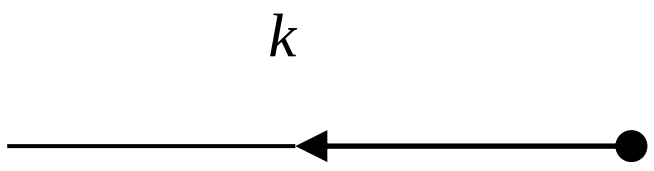

$\bar{v}_\sigma(k)$

Fig. 5. Antiparticle outgoing line.

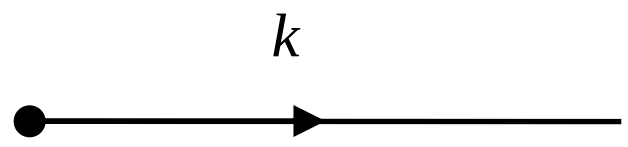

$v_\sigma(k)$

Anti-Dirac spinors

Fig. 6. Particle incoming line.

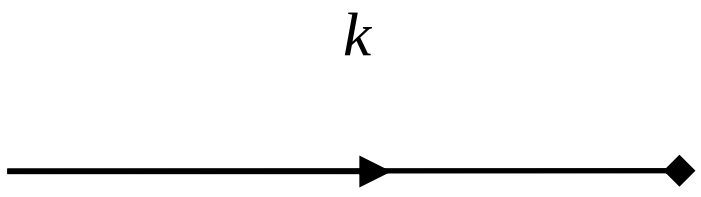

$\bar{u}_\sigma(-k)$

Fig. 7. Particle outgoing line.

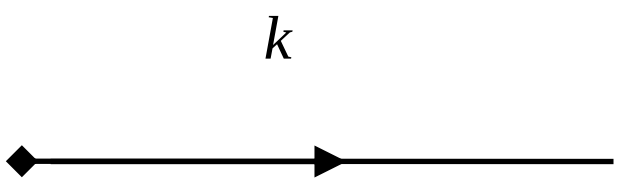

$u_\sigma(-k)$

Fig. 8. Antiparticle incoming line.

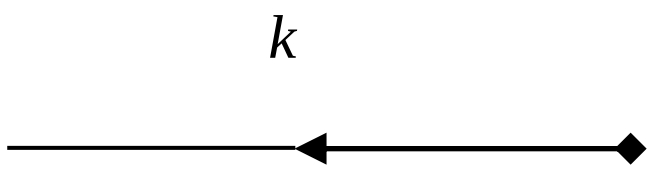

$v_\sigma(-k)$

Fig.9. Antiparticle outgoing line.

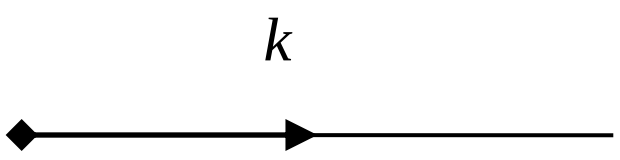

$\bar{v}_\sigma(-k)$

## Dirac-anti-Dirac spinors

Fig. 10. Particle incoming line.

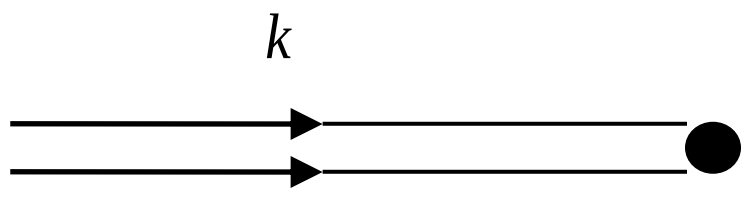

$u_{sp}^{A}(k)$

Fig. 11. Particle outgoing line.

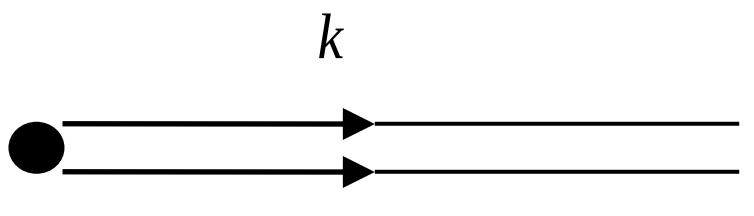

$\bar{u}_{sp}^{A}(k)$

Fig. 12. Antiparticle incoming line.

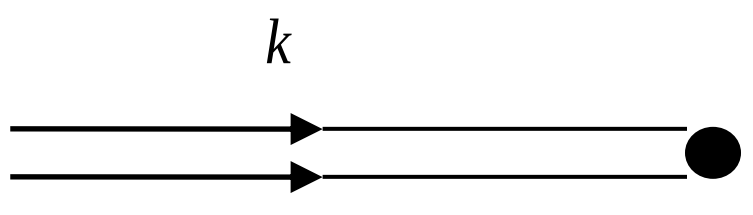

$v_{sp}^{A}(k)$

Fig. 13. Antiparticle outgoing line.

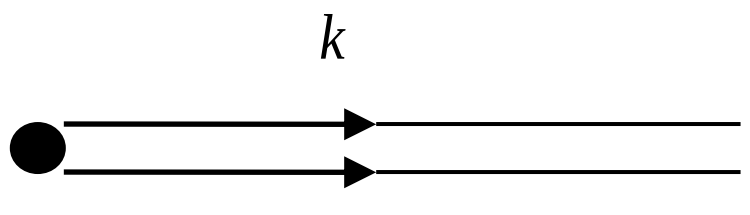

$\bar{v}_{sp}^{A}(k)$

## Propagators

Fig. 14. Scalar spin zero Dirac particle.

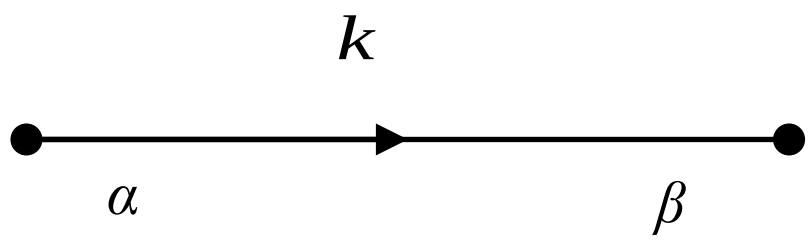

$$\frac{i\,(\not{k}+m)}{(k^2-m^2)+i\varepsilon}$$

Fig. 15. Scalar spin zero anti-Dirac particle.

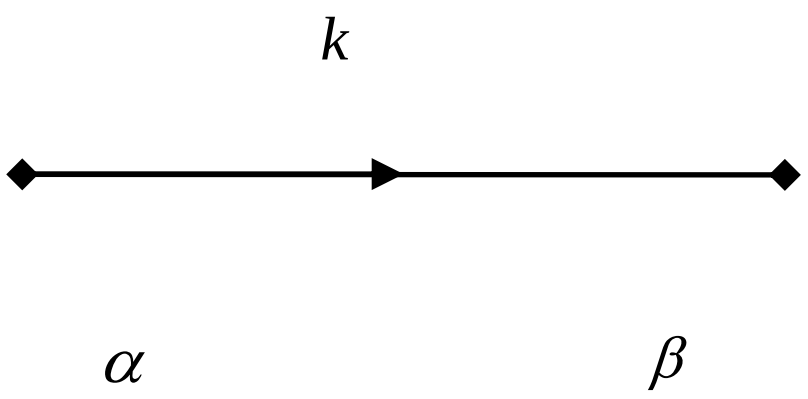

$$\frac{i\,(\not{k}+m)}{-(k^2-m^2)+i\varepsilon}$$

Fig. 16. Scalar spin one-half DaD doublet.

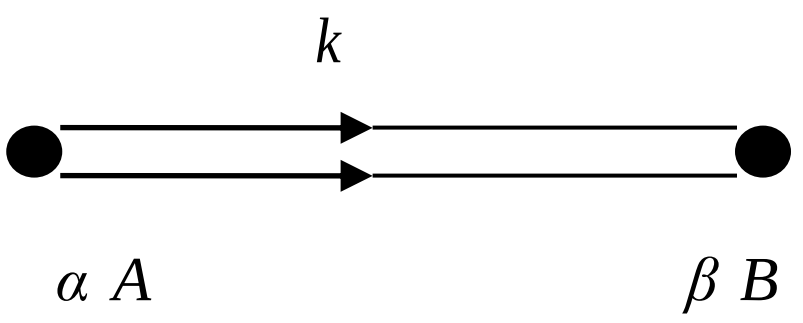

$$i\left((\not{k}-mc_{\lambda})\Gamma_3^{AB}-ims_{\lambda}\,\Gamma_1^{AB}\gamma^5+i\varepsilon\,\delta^{AB}\right)^{-1}$$

## Interaction Vertices

Fig. 17. Dirac vertex.

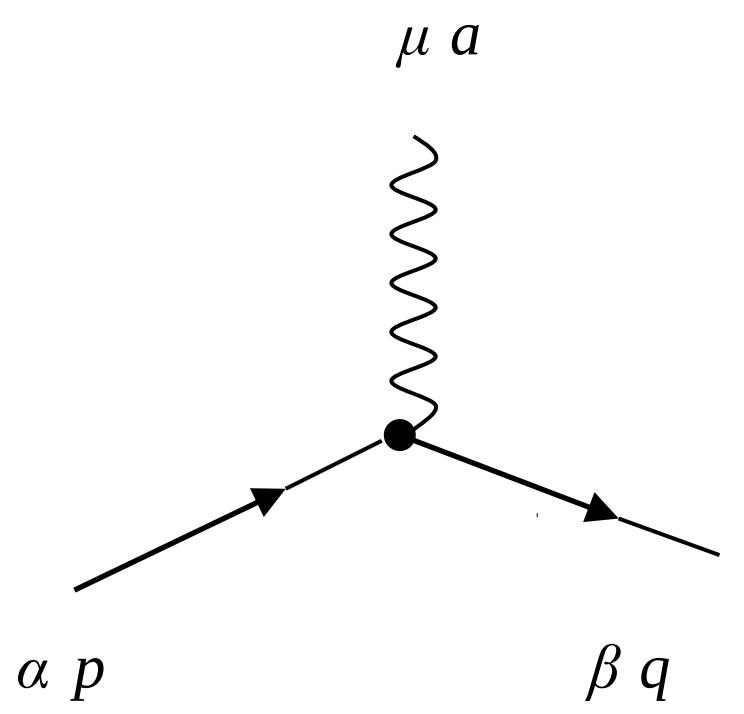

$$g\left(\gamma^{\mu}\right)_{\alpha\beta}\tau^{a}_{pq}$$

Fig. 18. Anti-Dirac vertex.

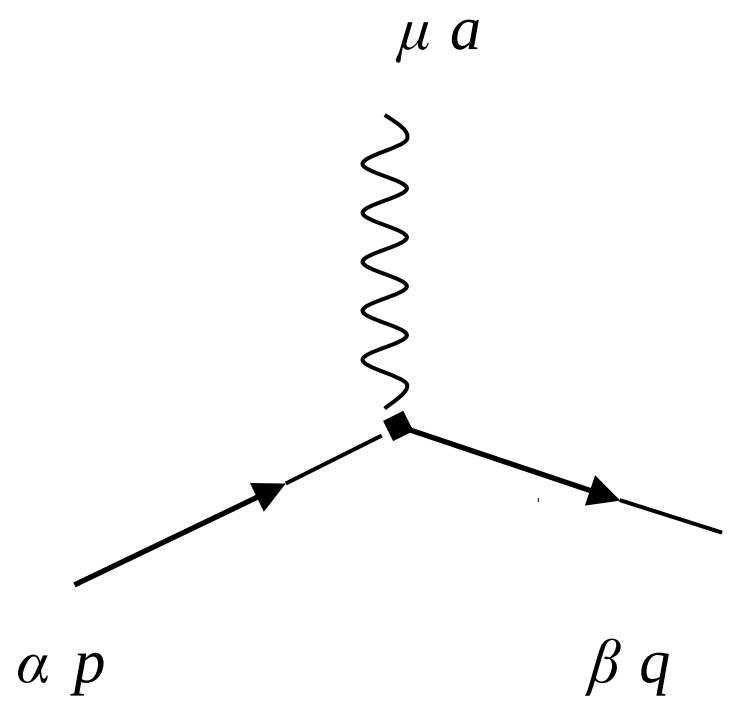

$$-g\left(\gamma^{\mu}\right)_{\alpha\beta}\tau^{a}_{pq}$$

Fig. 19. Dirac-anti-Dirac vertex.

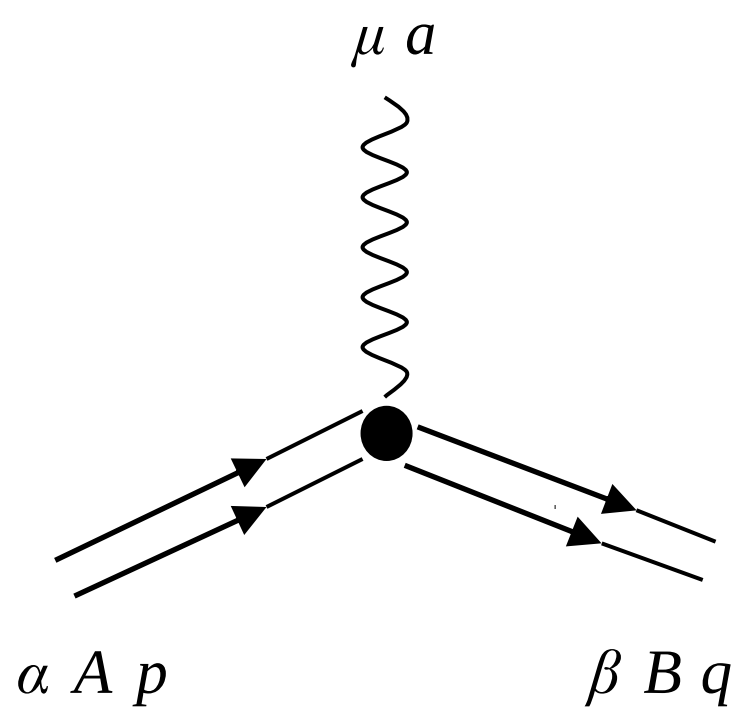

$$g\left(\Gamma_{3}\right)^{AB}\left(\gamma^{\mu}\right)_{\alpha\beta}\tau^{a}_{pq}$$